\documentclass[%
 reprint,
superscriptaddress,
nobibnotes,
 amsmath,amssymb,
science,
]{revtex4-2}
\usepackage{bibunits}
\usepackage{titletoc}
\usepackage{ragged2e}
\usepackage{graphicx}% Include figure files
\usepackage{dcolumn}% Align table columns on decimal point
\usepackage{bm}% bold math
\usepackage[colorlinks,
            linkcolor=black,
            citecolor=black,
            urlcolor=black
            ]{hyperref}
\usepackage{multirow}
\setcitestyle{round, numbers, italic}
\defaultbibliographystyle{sciencemag}
\defaultbibliography{reference}
\makeatletter
\let\auto@bib\@empty
\let\auto@bib@innerbib\@empty
\makeatother
\usepackage{float}
\usepackage{ragged2e}

\newcommand{\RepeatTitleBlock}{%
\begin{center}
{\Large\bfseries Supplementary Materials for Forecasting Return Time of Extreme Precipitation by Large Deviation Theory\par}
\vspace{1.2em}

{\normalsize
Haotian Xie$^{1,2,3}$, 
Haoxian Liu$^{1,2}$, 
Jingfang Fan$^{1,4}$, 
Ying Tang$^{5,6,7,8}$\par
}

\vspace{0.8em}

{\small
$^{1}$School of Systems Science, Beijing Normal University, Beijing 100875, China\par
$^{2}$Department of Systems Science, Faculty of Arts and Sciences, Beijing Normal University, Zhuhai 519087, China\par
$^{3}$Department of Industrial and Systems Engineering, The Hong Kong Polytechnic University, Hung Hom, Hong Kong\par
$^{4}$Potsdam Institute for Climate Impact Research, Potsdam 14412, Germany\par
$^{5}$Institute of Fundamental and Frontier Sciences, University of Electronic Science and Technology of China, Chengdu 611731, China\par
$^{6}$School of Physics, University of Electronic Science and Technology of China, Chengdu 611731, China\par
$^{7}$Key Laboratory of Quantum Physics and Photonic Quantum Information, Ministry of Education, University of Electronic Science and Technology of China, Chengdu 611731, China\par
$^{8}$Non-classical Information Science Basic Discipline Research Center of Sichuan Province, University of Electronic Science and Technology of China, Chengdu 611731, China\par
}

\vspace{0.6em}

{\small
These authors contributed equally.\par
Corresponding author: jamestang23@gmail.com\par
}
\end{center}
}

\begin{document}
\begin{bibunit}
%\preprint{APS/123-QED}
% \let\oldMakeTextUppercase\MakeTextUppercase
% \renewcommand{\MakeTextUppercase}[1]{#1}
\renewcommand{\thesubsection}{\arabic{section}.\arabic{subsection}}
\renewcommand{\thesection}{\arabic{section}}

\title{Forecasting Return Time of Extreme Precipitation by Large Deviation Theory}

\author{Haotian Xie}
\email[These authors contributed equally.]{}
\affiliation{School of Systems Science, Beijing Normal University, Beijing 100875, China}
\affiliation{Department of Systems Science, Faculty of Arts and Sciences, Beijing Normal University, Zhuhai 519087, China}
\affiliation{Department of Industrial and Systems Engineering, The Hong Kong Polytechnic University, Hung Hom, Hong Kong}

\author{Haoxian Liu}
\email[These authors contributed equally.]{}
\affiliation{School of Systems Science, Beijing Normal University, Beijing 100875, China}
\affiliation{Department of Systems Science, Faculty of Arts and Sciences, Beijing Normal University, Zhuhai 519087, China}

\author{Jingfang Fan}
\affiliation{School of Systems Science, Beijing Normal University, Beijing 100875, China}
\affiliation{Potsdam Institute for Climate Impact Research, Potsdam 14412, Germany}

\author{Ying Tang}
%\altaffiliation{These authors contributed equally}
\email[Corresponding authors: ]{jamestang23@gmail.com}
\affiliation{Institute of Fundamental and Frontier Sciences, University of Electronic Science and Technology of China, Chengdu 611731, China}
\affiliation{School of Physics, University of Electronic Science and Technology of China, Chengdu 611731, China}
\affiliation{Key Laboratory of Quantum Physics and Photonic Quantum Information, Ministry of Education, University of Electronic Science and Technology of China, Chengdu 611731, China}
\affiliation{Non-classical Information Science Basic Discipline Research Center of Sichuan Province, University of Electronic Science and Technology of China, Chengdu 611731, China}

\begin{abstract}
Forecasting extreme precipitation is essential yet challenging due to its rarity and complexity. We develop a large deviation framework to estimate the return times of extreme precipitation events. We first find that the Landau distribution, originally introduced in plasma physics, accurately captures extreme precipitation at approximately 93\% of global locations, outperforming conventional extreme value distributions with 76\% matched locations under the same accuracy criterion. Enriching rare event samples by the fitted Landau distribution, we obtain more accurate estimates of large deviation rate functions and return times, enabling forecasts beyond historically observed precipitation intensities. Mapping historical return times to future projections from the Coupled Model Intercomparison Project Phase 6 (CMIP6), we show that return time curves under different emission scenarios collapse onto a unified relation, revealing a sharply increased lifetime exposure to extreme precipitation for 21st-century birth cohorts under most future emission scenarios.
\end{abstract}

\maketitle

\section{Introduction} 

Extreme climate events, including heavy precipitation \cite{RevModPhys.92.035002,Donat2016508,Pfahl2017423}, heat waves \cite{Kim2023,Wanyama2023} and droughts \cite{Wang2023E943}, %have significant impacts on human societies and natural ecosystems 
profoundly impact societies and ecosystems~\cite{Kim2011175,Bell2018265,bao2024intensification}. Characterizing and forecasting extreme precipitation events is vital for understanding the critical phenomena of climate systems~\cite{fan2021statistical}. %, such as secondary disasters like floods \cite{Culler2021} and mudslides \cite{Lin20221951}. 
However, the distribution of extreme climate records such as precipitation intensities usually has a long tail encompassing rare but impactful events \cite{Zhong20241652}. The scarcity of recorded data for these infrequent climate events constrains the reliability of statistical methods \cite{Horton2016242,Marani2015121,Nerantzaki2019}, posing a significant challenge in forecasting extreme precipitation.

Many methods have been devised to characterize extreme climate events. Notably, statistical physics offers insights into extreme weather events by focusing on the underlying physical principles \cite{peters2001complexity,majumdar2019exactly, PhysRevLett.133.094201, lucarini2024detecting,liu2023teleconnections}. For precipitation, the seepage theory has improved the clustering of extreme precipitation \cite{Peters20092913}, and multifractal analysis has characterized scaling properties of precipitation fields \cite{Baranowski201539,Krzyszczak20191811}. Recently, machine learning algorithms \cite{RevModPhys.91.045002,Tang2022,choi2026advanced} such as  neural networks \cite{Madakumbura2021} have been employed to forecast time series. In precipitation modeling, neural general circulation models can reduce biases by leveraging observational data \cite{yuval2024neural}. The other approaches, such as causal inference models \cite{falasca2024data} and complex networks \cite{fan2021statistical,ludescher2021network,li2025comprehensive}, have identified extreme climate events. Despite tremendous efforts, due to the rarity of data, there is still a lack of a general approach to characterize the return times of extreme precipitation across the globe \cite{benestad2024global}.

\begin{figure*}[htbp]
%\centering
\includegraphics[width=\linewidth]{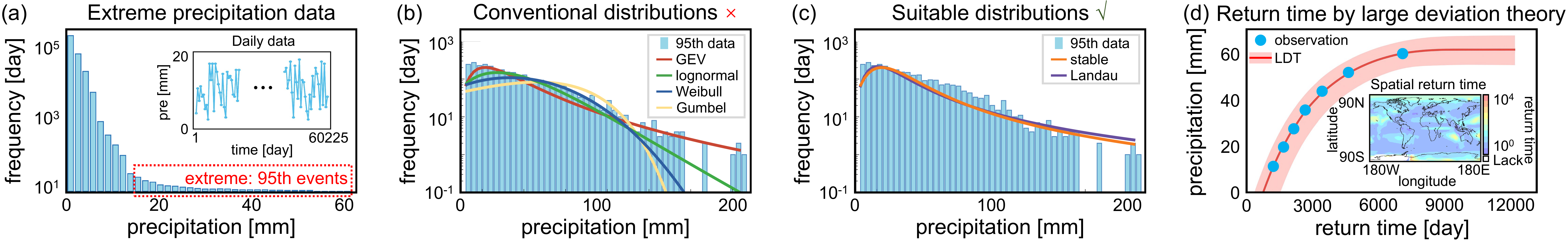}
\caption{\textbf{The large deviation framework for characterizing extreme precipitation and forecasting return time.} (a) The extreme precipitation events are taken as the daily 95th percentile of the precipitation in the Coupled Model Intercomparison Project Phase 6 (CMIP6) data \cite{li2021changes,saha2022rainfall} (Supplementary Figs.~S1-S2 ).
(b) The conventional fitting distributions have deviations in capturing the frequency of extreme precipitation. 
(c) Instead, the stable and Landau distributions are found to have higher accuracy. 
(d) The return time of precipitation at each location can be obtained from the large deviation theory. The inset shows the global view of return time for a fixed amount of precipitation.}
\label{fig.1}
%\vspace{-10pt}
\end{figure*}

In this paper, we develop an integrated framework for characterizing and forecasting extreme precipitation through the large deviation theory (LDT) \cite{touchette2009large,wilkinson2016large,Dematteis2018855,galfi2019large,PhysRevLett.127.058701,PhysRevLett.128.150603,tang_learning_2024} (Fig.~\ref{fig.1}). We find that the Landau distribution, known for its effectiveness in modeling heavy-tailed phenomena in plasma physics \cite{Horvathy2002The,Jacob2008Landau}, outperforms conventional distributions, such as Gumbel \cite{Qian2018212} and generalized extreme value (GEV) distributions \cite{Agilan201711,Sarhadi20172454}, in characterizing extreme precipitation, enabling more accurate sampling of precipitation tails. Inspired by the large deviation method on surface temperature \cite{PhysRevLett.127.058701}, we subsequently apply LDT to compute the return time of extreme precipitation. Using the rate function in the LDT, we obtain the return time of extreme precipitation.

We apply the approach to locations across the globe. Although precipitation distributions vary across locations, the Landau distribution fits 93\% of locations with a Hellinger distance below 0.3, outperforming other distributions by more than 15\% in accuracy, while requiring only two fitting parameters. Then, we employ the LDT to calculate the return time for given precipitation values. The predicted return times for extreme precipitation are consistent with the occurrence of historically recorded extreme events (Table~\ref{tab.em1}). To further evaluate the prediction, the quantile mapping method \cite{Enayati2021401} is employed to align scenario data from the CMIP6 model \cite{li2021changes,saha2022rainfall} under various emission scenarios. Using a single scaling factor, the return times from our framework and from the CMIP6 data are collapsed onto a unified relation, thereby enabling the prediction of return times under various emission scenarios and revealing increased lifetime exposure to extreme precipitation under most future emission scenarios.

\section{Results}
\subsection{Distributions of extreme precipitation} To predict the return time of extreme precipitation events, we first search for universal distributions of extreme precipitation. We use the Beijing Climate Center Earth System Model version 1 (BCC-ESM1) \cite{wu2020beijing} from CMIP6, excluding days with less than 0.1 mm of precipitation. We rank the remaining data and identify extreme precipitation as those beyond the 95th percentile \cite{camuffo2020relationship,mondal2020spatiotemporal} (Fig.~\ref{fig.1}a). We then fit the data of extreme precipitation to various distributions. Previous studies have commonly used the conventional extreme value distributions, including the Gumbel distribution \cite{osei2021estimation}, Weibull distribution \cite{olivera2019increases}, lognormal distribution \cite{papalexiou2013extreme}, and GEV distribution \cite{papalexiou2013battle}. Instead, we find that the stable distribution \cite{levy1925calcul} and the Landau distribution, a special case of the stable distribution used in plasma physics \cite{bulyak2022landau}, provide a better fit to the long tail of the distribution, leading to improved estimates of return times (Fig.~\ref{fig.1}b and \ref{fig.1}c).

\begin{figure*}[htbp]
%\centering
\includegraphics[width=0.65\linewidth]{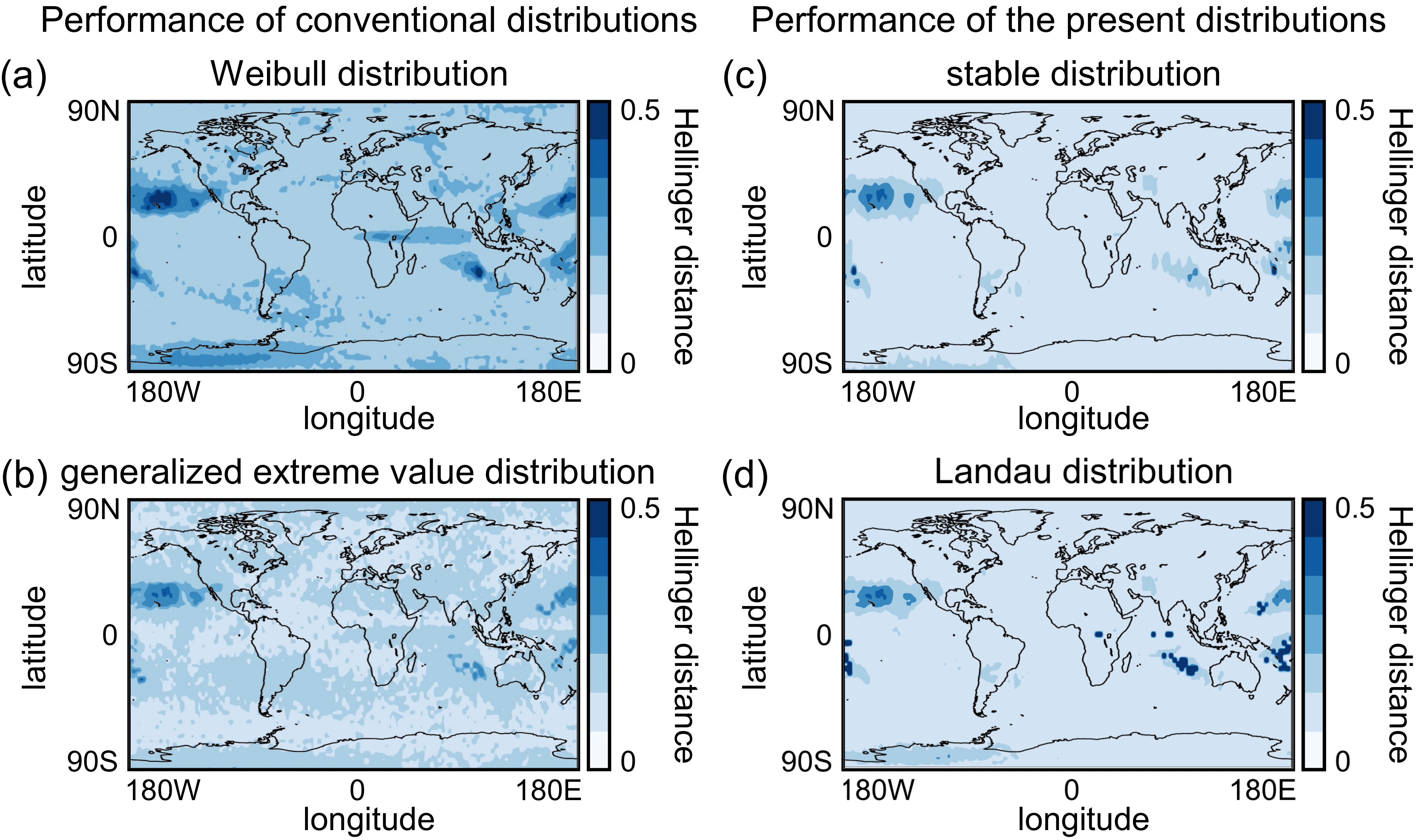}
\caption{\textbf{Performance of distributions in modeling extreme precipitation events.}
The (a) Weibull and (b) generalized extreme value distributions are used for modeling 95th percentile precipitation events and evaluated by Hellinger distance.
The (c) stable and (d) Landau distribution exhibit superior performance with smaller Hellinger distances.
}
%\vspace{-10pt}
\label{fig.2}
\end{figure*}

%We next describe the methodology and results of fitting extreme precipitation distributions. 
Specifically, to compare the fitted distribution $P$ with the historical data distribution $Q$, we use the Hellinger distance  
$H(P, Q) = \sqrt{\sum_{i=1}^m (\sqrt{p_i} - \sqrt{q_i})^2}/\sqrt{2}$ \cite{gonzalez2013class}, which ranges from 0 (perfect match) to 1 (complete divergence). %This metric evaluates the effectiveness of the fitted $P$ in capturing the observed historical precipitation data $Q$. 
This metric is particularly advantageous for heavy-tailed distributions as it is sensitive to discrepancies in low-probability density regions without being dominated by outliers. Other metrics such as the Jensen-Shannon divergence and the Kullback-Leibler divergence, show the same pattern (Supplementary Figs.~S3 ). In Fig.~\ref{fig.2}a and \ref{fig.2}b, the Hellinger distances for the Weibull and GEV distributions range from 0.25 to 0.5, indicating their insufficiency for accurate fitting.

%Thus, we seek for a more accurate distribution and find stable distribution suitable. 
Compared to the four conventional distributions, the Hellinger distance for the stable distribution is more stable and lower across most global regions (Fig.~\ref{fig.2}c). For $\alpha \neq 1$, the stable distribution $S(\alpha, \beta, \mu, c)$ \cite{levy1925calcul} is defined by its characteristic function
\begin{align}
\phi(t) = \exp \left\{ itc - |\mu t|^\alpha \left[ 1 - i\beta \operatorname{sign}(t) \tan \left( \frac{\pi \alpha}{2} \right) \right] \right\}, 
\end{align}
where $ 0 < \alpha \leq 2 $ is the stability parameter, $ -1 \leq \beta \leq 1 $ is the skewness parameter, $ \mu > 0 $ is the scale parameter, and $ c \in \mathbb{R} $ is the location parameter. At $\alpha=1$, %the tangent term becomes singular, and 
the characteristic function must be written separately in logarithmic form to preserve continuity.

We find that the Landau distribution $L(\mu, c)$, which is a special case of the stable distribution with $\alpha$ = 1 and $\beta$ = 1 \cite{bulyak2022landau}, leading to a simplified  characteristic function
\begin{align}
\phi(t) = \exp \left\{ itc - |\mu t| \left[ 1 + \frac{2i}{\pi} \operatorname{sign}(t) \ln |t| \right] \right\}. 
\end{align}
%The Landau distribution is a degenerate case of the stable distribution. 
Analogous to energy loss in plasma physics, the Landau distribution emerges as a statistical attractor for aggregated moisture subprocesses under the generalized central limit theorem, capturing the competition between background fluctuations and intense dynamical shocks. Here, the Landau distribution can accurately represent extreme precipitation (Fig.~\ref{fig.2}d). The results for the representative locations (Supplementary Figs.~S4-S9 ) show that both the stable
distribution and the Landau distribution perform well, with the Landau distribution requiring fewer fitting parameters.

\subsection{Parametric patterns in the fitted Landau distribution} 
\begin{figure*}[htbp]
%\centering
\includegraphics[width=0.65\linewidth]{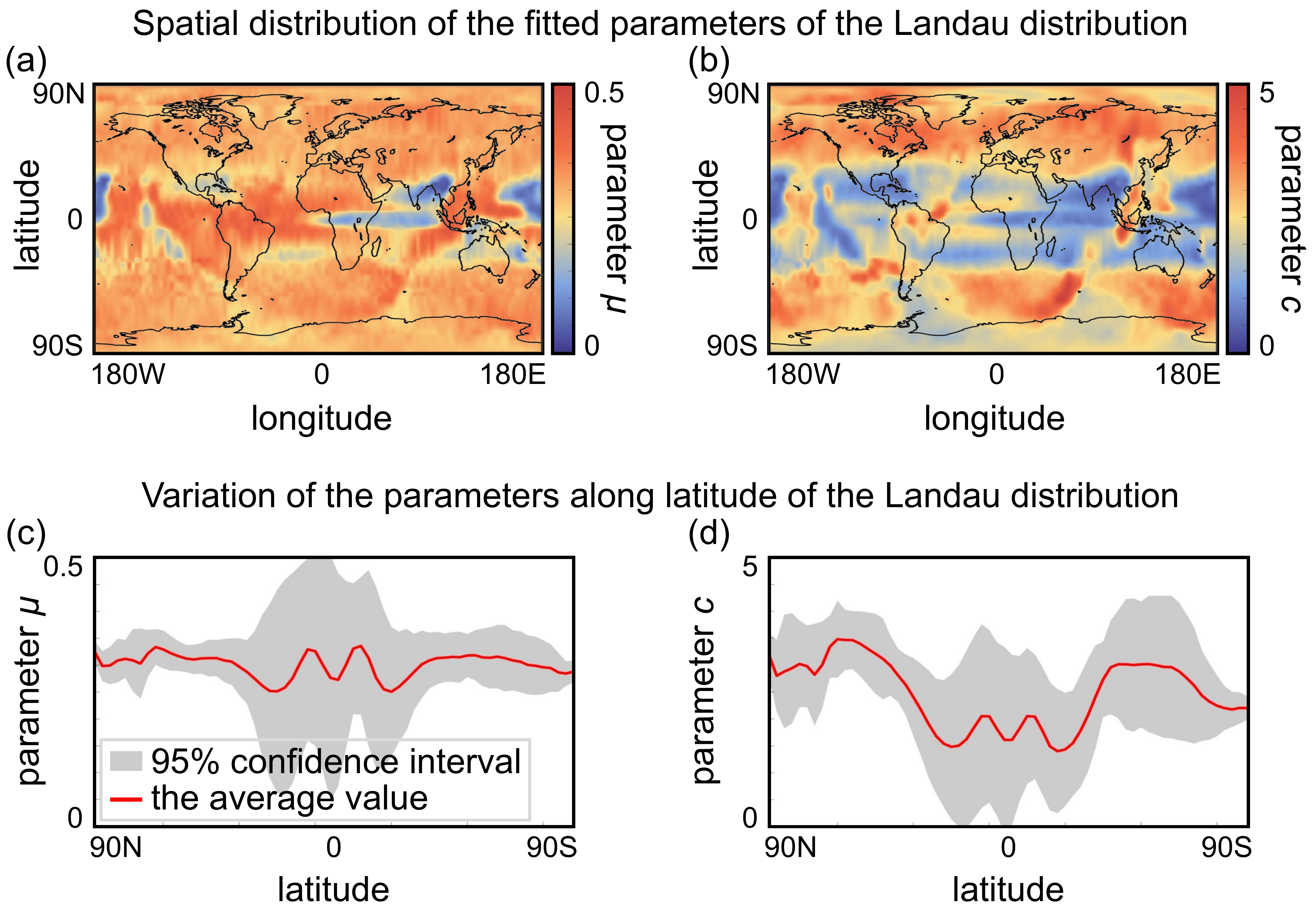}
\caption{\textbf{Global patterns of extreme precipitation distribution parameters.}
The values of $ \mu $ (a) and $ c $ (b) in the Landau distribution. The average (red line) and 95th percentile confidence intervals (gray shading) of the parameters $ \mu $ (c) and $ c $ (d) along latitude are shown.}
\label{fig.em1}
%\vspace{-10pt}
\end{figure*}

The fitting parameters of distributions suggest potential patterns across different locations. For example, $\mu$ and $c$ of the Landau distribution are smaller in low-latitude areas (Fig. \ref{fig.em1}a and \ref{fig.em1}b), indicating steeper and left-skewed fitted distributions, which may be linked to higher precipitation levels (Supplementary Figs.~S23 ). Figs.~\ref{fig.em1}c and ~\ref{fig.em1}d illustrate the trends of $\mu$ and $c$ for the Landau distribution along latitude with confidence intervals. Variations in latitude show two significant jumps around 60$^{\circ}$ and near low latitudes. These jumps suggest steeper and left-skewed distribution peaks, implying higher overall precipitation and increased probability of extreme precipitation in these regions. The patterns observed in the fitting parameters underscore the influence of geographical factors, such as latitude and proximity to the equator, on precipitation distribution.

\subsection{Return time} We next apply LDT \cite{wilkinson2016large,PhysRevLett.127.058701,galfi2019large} %, which was applied to study extreme climate events \cite{wilkinson2016large,PhysRevLett.127.058701,galfi2019large}, 
to capture the full spectrum of precipitation. %The correlation between precipitation and its frequency is discrete, which is absent in some extreme parts of the distribution. 
%The LDT helps estimate the probability of deviations for the extreme climate events \cite{wilkinson2016large,PhysRevLett.127.058701,galfi2019large}. 
Formally, let $ A_n=(1/n)\sum_{i=1}^nX_i $, where $X_i$ are identically distributed random variables. If in the limit of $n \to \infty$, there exists the rate function such that, for upper-tail events $a>a^*$,
\begin{equation}
I(a) = - \lim_{n \to \infty} \frac{1}{n} \ln P(A_n \ge a),
\end{equation}
then the exceedance probability scales as $P(A_n \ge a)\simeq\exp[-n I(a)]$. Here, $a^*$ is the typical value of $A_n$, satisfying $I(a^*)=0$, and $A_n \to a^*$ in probability as $n \to \infty$.
The corresponding return time for block length $n$ is
\begin{equation}
R_n(a)\simeq [P(A_n \ge a)]^{-1}\simeq\exp[n I(a)].
\end{equation}
When the rate function is approximately quadratic near $a^*$, $I(a)\approx (a-a^*)^{2}/\bigl(2\tau\sigma^{2}\bigr)$, $R_n(a)$ can be computed from the sample variance $\sigma^{2}$ and the integrated autocorrelation time $\tau$ of the recorded data~\cite{PhysRevLett.127.058701}.

\begin{figure*}[htbp]
%\centering
\includegraphics[width=0.65\linewidth]{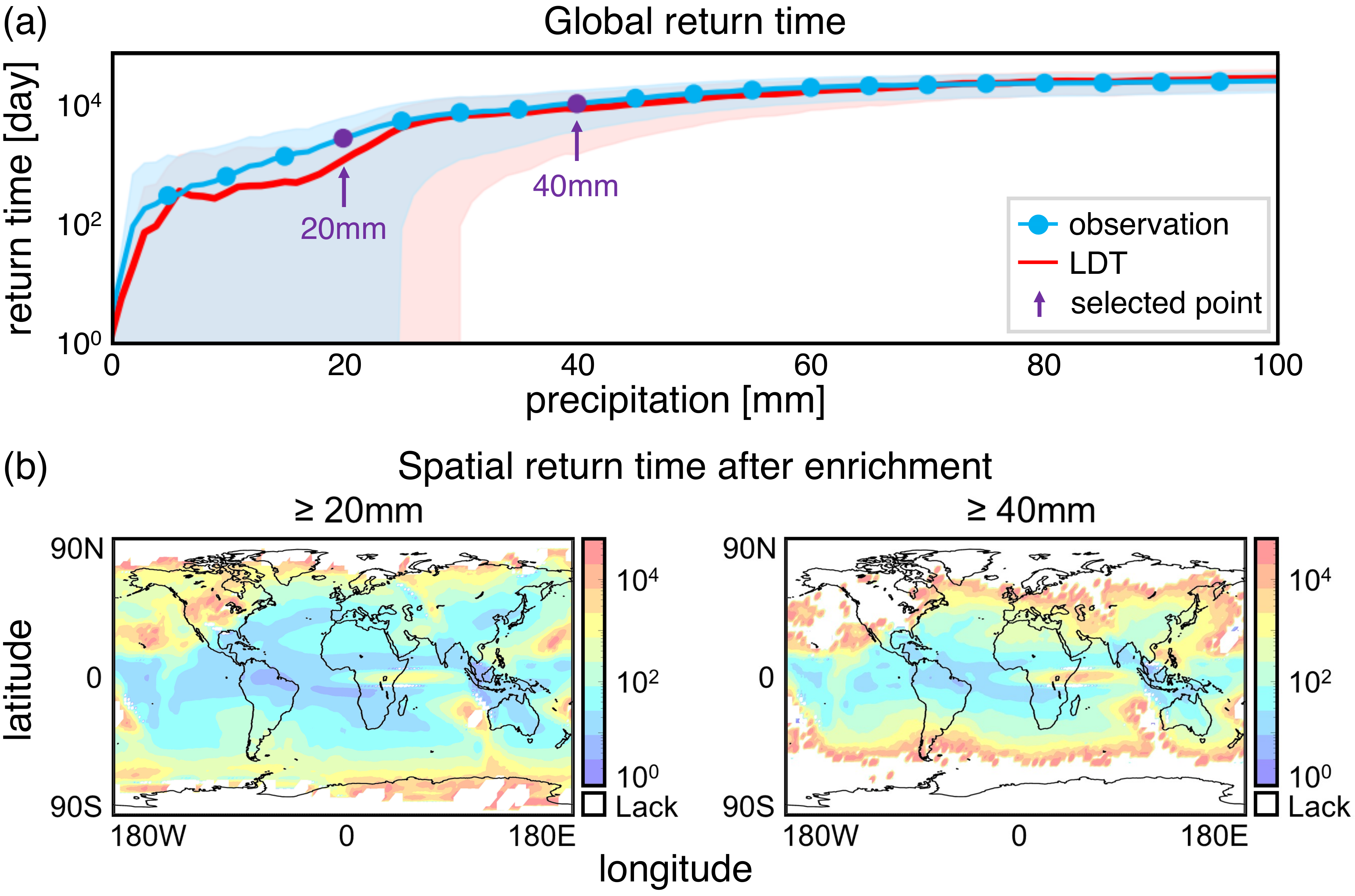}
\caption{\textbf{Precipitation return time across the globe based on the LDT.}
(a) Global precipitation return times are calculated using LDT, illustrating historical data (blue dots line) alongside theoretical predictions (red line). 
(b) The distribution of return times for precipitation thresholds of $\ge$ 20 mm (left) and $\ge$ 40 mm (right). Here, the fitted Landau distributions were used to sample and enrich the dataset to fill observational gaps for regions with rare precipitation.}
\label{fig.3}
%\vspace{5pt}
\end{figure*}

Based on this formulation, we determine the return time for varying precipitation values at locations across the globe (Figs.~\ref{fig.3}a,~\ref{fig.3}b). The return time curves derived from LDT effectively identify global precipitation trends. The difference in return time between various locations increases with higher precipitation levels. Low latitude locations have shorter return times, and this pattern extends poleward in the heat map. Central and northern North America exhibit fewer extreme events, with lower precipitation amounts on most wet days.

In areas with few extreme precipitation events, missing data reduces accuracy. To address this, we use the fitted Landau distribution to generate additional data for enrichment. Fig.~\ref{fig.3}b shows the results of this process, illustrating the before and after enrichment at the same precipitation level (Supplementary Figs.~S16-S18 ). At 40 mm, missing values in mid-latitudes, particularly oceanic regions, are adequately replenished. In particular, missing values in eastern North America, north-central Asia and Europe are enriched, which helps identify potential extreme precipitation events in these densely populated regions.

\begin{figure*}[htbp]
%\centering
\includegraphics[width=0.65\linewidth]{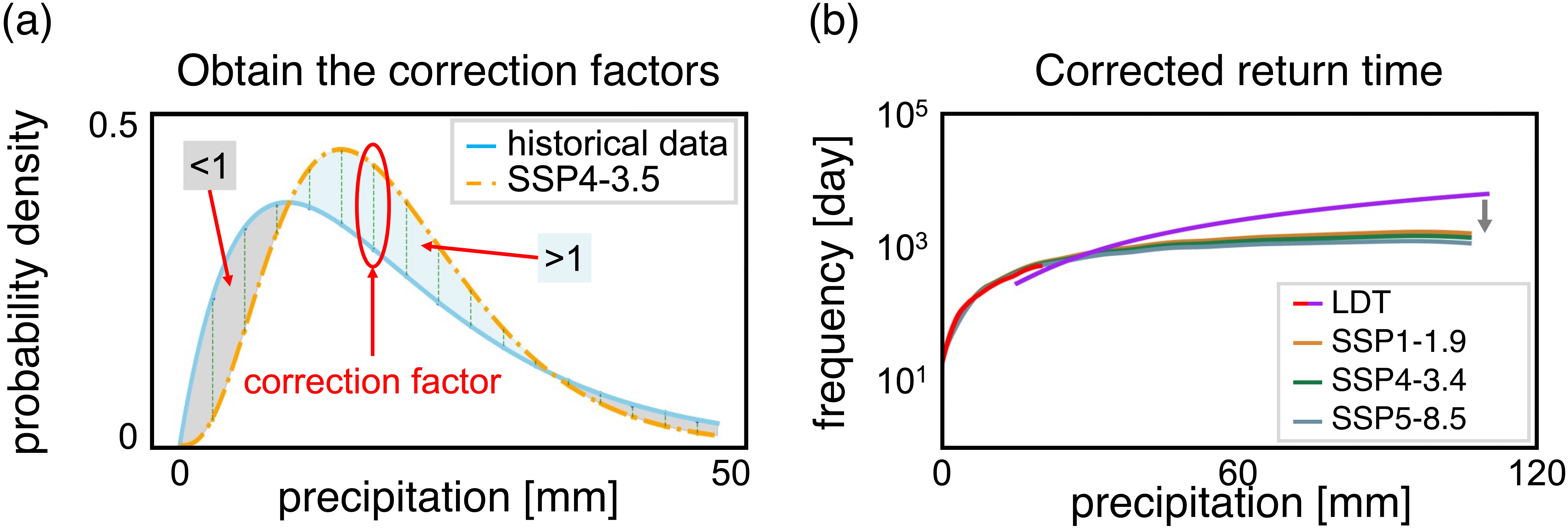}
\caption{\textbf{Projected return time of future precipitation under various emission scenarios. } (a) Corrected precipitation return time projections using quantile mapping. Return times from historical and scenario data are matched, and their ratios give correction factors (red circles).
(b) Corrected return time curve for Jakarta as an example. For the return time, the red denotes CMIP6 data, and the purple denotes LDT predictions. Grey arrows denote shifts from historical to future values. Similar patterns hold in other locations (Supplementary Figs.~S19-S22 ).}
\label{fig.4}
%\vspace{-19pt}
\end{figure*}

\subsection{Return time for representative cities} To validate the effectiveness of our method, we assess the return times for 10 populous cities \cite{de2012vulnerability} (Supplementary Tables ~S1, and S2  for more cities), distributed across East and South Asia, Africa, North America, and Latin America. Using the Landau distribution, the predicted return times are compared to historical data as shown in Table~\ref{tab.em1}. Generally, the predictions for extreme precipitation levels (e.g., 70 mm) maintain high accuracy. In cities with high average precipitation, such as Jakarta and Kinshasa, the regression frequency bias for a 30-year return period of 70 mm is minimal, with an error of less than 10. This demonstrates the effectiveness of the Landau distribution in capturing and identifying extreme precipitation events and the predictability by the LDT method.

%\vfill\break 
\begin{table}[htbp]
\caption{\textbf{Predicted frequencies of extreme precipitation and corresponding historical counts for 10 populous cities.} For each city, the second row reports the predicted number of extreme precipitation events, and the first row reports the corresponding historical count. The precipitation thresholds are 40 mm and 70 mm over 10-, 20-, and 30-year periods.}
\begin{ruledtabular}
\begin{tabular}{lllllll}
\multirow{2}{*}{city}         & \multicolumn{2}{c}{10-year} & \multicolumn{2}{c}{20-year} & \multicolumn{2}{c}{30-year} \\ \cline{2-7} 
                              & 40 mm         & 70 mm         & 40 mm         & 70 mm         & 40 mm         & 70 mm         \\ \hline
\multirow{2}{*}{Beijing}      & 20           & 1            & 40           & 3            & 60           & 5            \\
                              & 22           & 1            & 44           & 4            & 66           & 5            \\
\multirow{2}{*}{Jakarta}      & 102          & 29           & 203          & 58           & 305          & 87           \\
                              & 111          & 32           & 221          & 63           & 332          & 95           \\
\multirow{2}{*}{Kinshasa}     & 99           & 31           & 198          & 62           & 298          & 93           \\
                              & 108          & 34           & 216          & 68           & 324          & 101          \\
\multirow{2}{*}{Lagos}        & 47           & 11           & 93           & 22           & 140          & 33           \\
                              & 51           & 12           & 102          & 24           & 153          & 36           \\
\multirow{2}{*}{Manila}       & 186          & 77           & 371          & 154          & 557          & 231          \\
                              & 202          & 84           & 404          & 168          & 606          & 251          \\
\multirow{2}{*}{Mexico City}  & 173          & 71           & 346          & 142          & 518          & 212          \\
                              & 188          & 77           & 376          & 154          & 564          & 231          \\
\multirow{2}{*}{New York}     & 30           & 2            & 60           & 5            & 90           & 7            \\
                              & 33           & 2            & 65           & 5            & 98           & 7            \\
\multirow{2}{*}{Osaka}        & 63           & 10           & 127          & 21           & 190          & 31           \\
                              & 69           & 11           & 138          & 23           & 207          & 34           \\
\multirow{2}{*}{Sao Paulo}    & 3            & 1            & 6            & 1            & 9            & 2            \\
                              & 3            & 1            & 6            & 1            & 9            & 2            \\
\multirow{2}{*}{Tokyo}        & 51           & 7            & 102          & 13           & 153          & 20           \\
                              & 56           & 7            & 111          & 15           & 167          & 22          
\end{tabular}
\end{ruledtabular}
\label{tab.em1}
\end{table}

\subsection{Forecasting future precipitation with model validation} Analyzing coming shifts in precipitation becomes essential for urban planning under climate change. Thus, we compare the predicted return time with the CMIP6 prospective data \cite{li2021changes,saha2022rainfall,li2025constraining} under various emission scenarios. To this end, we use a quantile mapping method \cite{Enayati2021401} to align return time projections by applying a correction factor. This factor is derived from matching precipitation distributions of historical and CMIP6 data  (Fig.~\ref{fig.4}a), and % precipitation distributions from CMIP6 data under three climate scenarios are compared to historical data to derive adjustments that align their return time with the LDT predictions. 
then used to align the return time of the LDT prediction and CMIP6 data. After applying the correction factor, the forecasted return times under all climate scenarios collapse to a unified relation (Fig.~\ref{fig.4}b), showing that return time projections are consistent between the LDT prediction and CMIP6 prospective data. We note that only a single correction factor is needed to convert the return-time curves of different scenarios to collapse on the unified relation.
% different climate scenarios only need a correction factor to collapse on the same relation of return time.

\begin{figure*}[htbp]
%\centering
\includegraphics[width=0.65\linewidth]{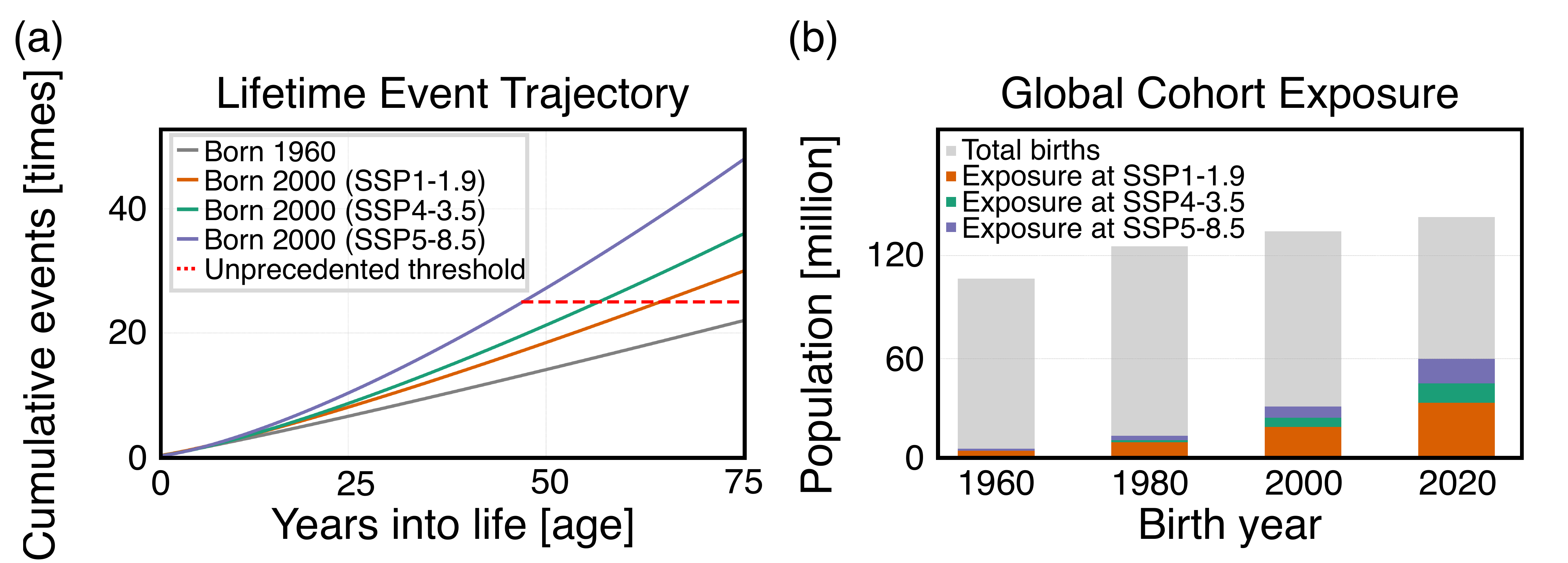}
\caption{\textbf{Predicted lifetime exposure to extreme precipitation.} (a) The lifetime accumulation of extreme precipitation events is projected to accelerate sharply for younger generations that may cross unprecedented exposure thresholds at mid-life (dashed red line)~\cite{grant2025global}. (b) A rapidly growing portion of the population is projected to exceed the unprecedented exposure threshold.}
\label{fig.5}
%\vspace{-13pt}
\end{figure*}

We further explore how lifetime exposure \cite{grant2025global} to extreme precipitation will evolve for different birth cohorts under various emission scenarios. Fig.~\ref{fig.5}a contrasts the accelerating lifetime risk trajectories for different generations against an unprecedented exposure threshold, while Fig.~\ref{fig.5}b quantifies the absolute number of people within successive global birth cohorts projected to surpass this risk level. Not only is the lifelong event frequency for any given individual projected to increase, but the total global population subjected to this high-risk regime is also set to expand. Thus, the disproportionate burden of future climate change is placed squarely on younger generations, who will experience extreme precipitation more severe than their predecessors. 

\section{Discussion} We have developed an integrated framework for identifying and forecasting extreme precipitation. Precipitation values fluctuate from one location to another and are contingent upon disparate geographic conditions, causing a challenge to have uniform quantitative rules that can capture these diverse characteristics \cite{knapp2008consequences,vazquez2017ecological,zumwald2020understanding}. Despite the challenges, we make an advance by discovering the Landau distribution~\cite{papalexiou2013extreme,nerantzaki2019tails} and employing LDT to characterize such data. Specifically, the heavy-tailed characteristics of precipitation are effectively captured by the Landau distribution, where heavy precipitation can be viewed as the aggregation of numerous weakly correlated subprocesses, analogous to Landau's original formulation for energy loss in high-speed particles~\cite{breugem2020meteorological}. %The suitability of Landau distribution enables to characterize the extreme precipitation across the globe including the areas of low precipitation, and then predict return time of that did not occur in the past.  
The suitability of the Landau distribution enables us to characterize extreme precipitation across the globe, including regions with low precipitation, and to predict return times for events that have not occurred in the historical record.

The selection of the raw data requires a careful treatment to reach a consistent fitting process of distributions. For instance, the fitted distribution is influenced by the percentile of precipitation days selected and categorized \cite{schar2016percentile}. In our case, the 95th percentile and removal of precipitation days less than 0.1 mm were selected in a way that maximizes the retention of extreme precipitation values while also avoiding errors associated with small precipitation \cite{camuffo2020relationship,mondal2020spatiotemporal} (Supplementary Figs.~S4-S9 ). When calculating the enriched return time, we also compared the differences introduced by different choices of percentile and precipitation days, and found that the selections did not affect the range or trend of the return time, demonstrating the robustness of the results (Supplementary Figs.~S13-S15 ).

In the calculation of the rate function, the average block length $n$ can affect the estimate of the return time. The use of larger blocks may result in the dilution of extreme events and an underestimation of the rate function \cite{PhysRevLett.127.058701}. We have compared the impact of varying $n$ (Supplementary Figs.~S10-S12 ). We found that at $n = 1$ the statistical properties of individual precipitation events can be preserved without averaging effects, thus avoiding the smoothing effect of larger blocks that mask extreme events. Under this setting, the return time can be calculated for a greater number of locations and obtained for more extreme precipitation values.

In this study, we focus on daily extremes to derive return time relations at multi-decadal scales. The fitted extreme precipitation distributions allow for the assessment of precipitation in different regions and the generation of stable quantitative forecasts. Our framework can also be applied to sub-daily data, but such an extension requires a redefinition of the parameters and thresholds mentioned above. As a potential avenue for improvement, the distribution parameters can be estimated using methods such as metaheuristic optimization algorithms \cite{alrashidi2020metaheuristic}. Besides, the incorporation of other climatic variables and a more detailed spatial resolution could enhance the predictive capability of our framework \cite{xu2021spatiotemporal,gettelman2022future}. Thus, our results highlight the potential of LDT for revealing universal patterns of extreme climate events.

% The fitted extreme precipitation distributions allow for the assessment of precipitation in different regions and the generation of stable quantitative forecasts. As a potential avenue for improvement, the distribution parameters can be estimated using methods such as metaheuristic optimization algorithms \cite{alrashidi2020metaheuristic}. Besides, the incorporation of other climatic variables and a more detailed spatial resolution could enhance the predictive capability of our framework \cite{xu2021spatiotemporal,gettelman2022future}. Thus, our results highlight the potential of LDT on revealing universal patterns of extreme climate events.

\section*{Materials and Methods}

%\section{Methodology}

\subsection{Distributions for Extreme Precipitation}

Accurate modeling of precipitation distributions can help us understand the underlying statistical behavior of extreme precipitation and predict future occurrences. Several statistical distributions are commonly used to fit precipitation data, each offering different characteristics that may suit various data types. This section explores how the Gumbel, Weibull, logarithmic normal (lognormal), generalized extreme value (GEV), stable, and Landau distributions can be applied to model the distribution of precipitation events.

\subsubsection{Gumbel Distribution}

The Gumbel distribution, also known as the extreme value type I distribution, is frequently used in hydrology and meteorology to model the maximum (or minimum) distribution of a sample of independent and identically distributed random variables ~\cite{gumbel1958statistics,gnedenko1943distribution}. The probability density function (PDF) of the Gumbel distribution is given by
\begin{equation}
f(x) = \frac{1}{\beta} \exp\left(-\frac{x - \mu}{\beta}\right) \exp\left(-\exp\left(-\frac{x - \mu}{\beta}\right)\right),
\end{equation}
where $ \mu $ is the location parameter, and $ \beta $ is the scale parameter.

\subsubsection{Weibull Distribution}

The Weibull distribution is another flexible distribution often used in reliability analysis and can be applied to model precipitation data ~\cite{nelson2003applied}. The PDF is given by
\begin{equation}
f(x) = \frac{k}{\lambda}\left(\frac{x}{\lambda}\right)^{k-1}\exp\left(-\left(\frac{x}{\lambda}\right)^k\right),
\end{equation}
where $ k $ is the shape parameter and $ \lambda $ is the scale parameter.

\subsubsection{Logarithmic Normal Distribution}

The lognormal distribution is suitable for positively skewed data such as precipitation amounts ~\cite{gaddum1945lognormal}. If $\log(x)$ follows a normal distribution, then $x$ is said to follow a lognormal distribution. The PDF is
\begin{equation}
f(x) = \frac{1}{x \sigma \sqrt{2\pi}} \exp\left(-\frac{(\log(x) - \mu)^2}{2\sigma^2}\right),
\end{equation}
where $ \mu $ and $ \sigma $ are the mean and standard deviation of the logarithm of the variable, respectively.

\subsubsection{Generalized Extreme Value Distribution}

The GEV distribution unifies the Gumbel, Fr\'echet, and Weibull distributions into a single family and is used to model the distribution of extreme events ~\cite{von1936distribution,jenkinson1955frequency}. It has three parameters, location $ \mu $, scale $ \sigma $, and shape $ \xi $, and its cumulative distribution function (CDF) is expressed as follows
\begin{equation}
F(x) = \exp\left[-\left(1 + \xi\frac{x - \mu}{\sigma}\right)^{-1/\xi}\right],
\end{equation}
where $ 1 + \xi\frac{x - \mu}{\sigma} > 0 $.

\subsubsection{Stable Distribution}

The stable distribution, also known as the L\'evy alpha-stable distribution, is a generalization of the normal distribution that allows for heavy tails and skewness. Unlike the normal distribution, the stable distribution has no closed-form PDF except in a few special cases ~\cite{levy1925calcul}. It is characterized by four parameters: the stability parameter $ \alpha $ with $0< \alpha \le 2$, the skewness parameter $ \beta $ with $-1\le \beta \le 1$, the scale parameter $ \mu > 0$, and the location parameter $ c $. For $ \alpha \neq 1 $, the characteristic function is given by
\begin{equation}
\phi(t) = \exp\left\{ itc - \mu |t|^\alpha \left[ 1 - i\beta \,\mathrm{sign}(t)\tan\left(\frac{\pi\alpha}{2}\right) \right] \right\},
\end{equation}
where $ \mathrm{sign}(t) $ is the sign function.

\subsubsection{Landau Distribution}

The Landau distribution is another heavy-tailed distribution, originally derived in quantum field theory, that has found applications in modeling extreme events in various fields ~\cite{landau1944energy,bulyak2022landau}. Like the stable distribution, the Landau distribution has a heavy tail, making it suitable for capturing extreme precipitation events. It can be regarded as a special case of the stable distribution and does not admit a simple elementary closed-form PDF. We therefore denote its PDF by
\begin{equation}
f(x)=f_{\mathrm{L}}(x;c,\mu),
\end{equation}
where $ c $ and $ \mu $ are the location and scale parameters, respectively.

\subsection{Large Deviation Analysis of Time-averaged Variables}

We explore the large deviations of time averages of observables in dynamical systems, taking the averaging time $T$ as the large deviation parameter. Consider a Markov process $x(t) \in \mathbb{R}^n$ and an observable $A: \mathbb{R}^n \rightarrow \mathbb{R}$. The time average
\begin{equation}
a=\frac{1}{T} \int_0^T A(x(t)) \mathrm{d} t
\end{equation}
has density
\begin{equation}
\rho(a) \underset{T \rightarrow \infty}{\asymp} \mathrm{e}^{-T I(a)},
\end{equation}
where $I(a)$ is the rate function. It is often convenient to introduce the scaled cumulant generating function
\begin{equation}
\lambda(k)=\lim _{T \rightarrow+\infty} \frac{1}{T} \log \mathbb{E}\left[\exp\left(k \int_0^T A(x(t)) \mathrm{d} t\right)\right].
\end{equation}
Under the conditions of the G\"artner-Ellis theorem~\cite{chen2000generalization}, $I(a)$ and $\lambda(k)$ are related through a Legendre-Fenchel transform. When $\lambda(k)$ is differentiable, the rate function is
\begin{equation}
I(a)=k(a)\, a-\lambda(k(a)),
\end{equation}
where $k(a)$ is determined by
\begin{equation}
a=\lambda^{\prime}(k(a)).
\end{equation}

Such a large deviation principle can hold for sufficiently mixing dynamics and suitable observables. For ergodic Markov processes, Donsker-Varadhan theory provides the relevant framework~\cite{donsker1976asymptotic}. In applications, however, whether the asymptotic regime has been reached must still be checked empirically. In systems with strong memory or long-range correlations, anomalous large-deviation scalings with powers different from $T$ may occur.

For ergodic systems, the time average $a$ converges to the ergodic average $\mu=\mathbb{E}[A]$ as $T$ increases. Under mixing assumptions, the central limit theorem implies
\begin{equation}
\sqrt{T}\,(a-\mu) \sim N\left(0, \sigma^2 \tau_c\right),
\end{equation}
where $\sigma^2$ is the variance of $A$, and $\tau_c$ is its integrated autocorrelation time.

Large deviation theory extends this Gaussian regime to rare fluctuations of the time average. Expanding the rate function around its minimum at $a=\mu$, one obtains
\begin{equation}
\rho(a) \asymp \mathrm{e}^{-T(a-\mu)^2 I^{\prime \prime}(\mu) / 2}.
\end{equation}
Accordingly, the Gaussian approximation for $a$ has variance $1/\bigl[T I^{\prime \prime}(\mu)\bigr]$. Equivalently, near $a=\mu$,
\begin{equation}
I(a)\approx \frac{(a-\mu)^2}{2\sigma^2 \tau_c}.
\end{equation}
In general, $I(a)$ contains information beyond Gaussian fluctuations, and its higher-order structure characterizes deviations from Gaussianity and the behavior of the tails.

For ergodic Markov processes, the large deviation function for additive observables can be studied within Donsker-Varadhan theory. However, in complex systems such as climate dynamics, it is often estimated empirically. One practical method is to divide the time series of $A(x(t))$ into blocks of length $\tau_b \gg \tau_c$. The time average over these blocks is
\begin{equation}
a_b^j=\frac{1}{\tau_b} \int_{(j-1) \tau_b}^{j \tau_b} A(x(t)) \mathrm{d} t.
\end{equation}
Under sufficient mixing and for large enough $\tau_b$, these block averages can be treated as approximately independent realizations of the time-averaged observable. For finite $\tau_b$, the corresponding rate function can be estimated by
\begin{equation}
I_b(a)=-\frac{\ln \rho(a)}{\tau_b},
\end{equation}
where $\rho(a)$ denotes the density of $a_b^j$. Its convergence can be assessed by increasing $\tau_b$ until the estimates stabilize within an acceptable error margin. Alternatively, the scaled cumulant generating function can be computed first, and the G\"artner-Ellis theorem can then be applied to obtain the rate function~\cite{rohwer2015convergence,ragone2020computation}. This method avoids direct estimation of the density and often requires less data for comparable precision.

Both methods often require extensive data to capture non-Gaussian fluctuations in climate science applications. The challenge lies in accurately estimating the tails of the large deviation function. In conclusion, a balance among the block length $\tau_b$, the number of blocks $N_b$, and the time scales associated with autocorrelation and mixing must be carefully managed to achieve accurate large deviation estimates.

With the rate function $ I(a) $ determined, we can proceed to calculate the return time $ T_R(a) $ for an exceedance event with threshold $ a $. For upper-tail events above the typical value $\mu$, the exceedance probability scales as
\begin{equation}
P\left(\frac{1}{T}\int_0^T A(x(t)) \mathrm{d} t \ge a\right) \asymp e^{-T I(a)}.
\end{equation}
Thus, the return time $ T_R(a) $ can be approximated by
\begin{equation}
T_R(a) \approx P\left(\frac{1}{T}\int_0^T A(x(t)) \mathrm{d} t \ge a\right)^{-1} \approx e^{T I(a)}.
\end{equation}
A smaller rate function indicates more frequent exceedance events, while a larger rate function corresponds to rarer occurrences.

\subsection{Data Enrichment by Sampling from Fitted Distribution}

Modeling the tails of the distribution is important for analyzing extreme precipitation events, especially when observations of the most extreme events are sparse or missing. Given the Landau distribution's ability to effectively model the heavy-tailed nature of precipitation data, it can be leveraged to sample the missing frequencies corresponding to extreme precipitation amounts. This process begins by fitting the Landau distribution to the historical data, particularly focusing on the tail where the extreme events lie. Let $ x $ denote the precipitation amount and $ f_L(x) $ represent the fitted Landau distribution's PDF. The CDF $ F_L(x) $ can be derived from $ f_L(x) $ and is given by
\begin{equation}
F_L(x) = \int_{-\infty}^x f_L(u) \, du.
\end{equation}
To sample the missing frequencies, particularly in the tail, we generate a series of random variables $ x_i $ that follow the Landau distribution
\begin{equation}
X_i \sim \mathrm{Landau}(c,\mu),
\end{equation}
where $ c $ and $ \mu $ are the location and scale parameters of the Landau distribution, respectively.

For each sampled value $ x_i $, the corresponding expected frequency over a small interval around $ x_i $ is approximated by
\begin{equation}
f(x_i) \approx n \cdot f_L(x_i)\,\Delta x,
\end{equation}
where $ n $ is the total number of historical precipitation events and $ \Delta x $ is the interval width. This approach allows us to estimate the frequencies of extreme events that are not observed in the empirical dataset, thereby filling gaps in the tail of the distribution.

\subsection{The Quantile Mapping Method}

The quantile mapping is a statistical technique widely used to correct biases in climate model outputs by aligning the modeled distribution of a climate variable with historical data. For extreme precipitation, quantile mapping can be used to adjust future climate scenarios using precipitation return times obtained from LDT and the Landau distribution. This section describes how quantile mapping is used to derive correction factors for different future climate scenarios and to adjust return times accordingly.

Let $ F_{\text{obs}}(x) $ be the empirical CDF of the historical precipitation amounts, and $ F_{\text{mod}}(x) $ be the CDF of the modeled precipitation amounts for a future climate scenario. The quantile mapping function $ QM(x) $ is defined as
\begin{equation}
QM(x) = F_{\text{obs}}^{-1}(F_{\text{mod}}(x)),
\end{equation}
where $ F_{\text{obs}}^{-1} $ is the inverse CDF of the historical data. This function maps the modeled precipitation amount $ x $ to the corresponding value in the historical distribution, effectively correcting the modeled data.

The correction factor $ CF(x) $ for a specific precipitation amount $ x $ is then given by
\begin{equation}
CF(x) = \frac{QM(x)}{x}.
\end{equation}
This factor represents the ratio of the historical precipitation amount to the modeled amount at the same quantile, providing a means to adjust the future scenario data.

Once the correction factors are derived, they are used to adjust precipitation amounts under different future climate scenarios rather than to rescale return times directly. The correction process therefore maps the precipitation amount associated with a given quantile to its corrected value under the future climate scenario and then evaluates the corresponding return time through the return-time relation obtained from large deviation theory.

Let $ T(x) $ denote the return time for a precipitation amount $ x $ under current climate conditions, as derived from large deviation theory. The mapped return time $ T_{\text{mapped}}(x) $ under a future climate scenario is then given by
\begin{equation}
T_{\text{mapped}}(x) = T(QM(x)),
\end{equation}
where $ QM(x) $ is the quantile-mapped precipitation amount. This adjustment accounts for changes in the precipitation distribution under the future scenario, as represented by the climate model, while preserving the return-time relation derived from large deviation theory.

For instance, if the correction factor implies that the mapped precipitation amount is increased by 10\%, then $ QM(x)=1.1x $. In that case, the mapped return time is evaluated at the corrected precipitation amount,
\begin{equation}
T_{\text{mapped}}(x) = T(1.1x).
\end{equation}
This process is repeated across the full range of precipitation amounts to generate a complete set of adjusted return times for the future scenario.

%\appendix

%\section{Appendixes}

\section*{Supplementary Materials}

\textbf{The PDF file includes:}

Supplementary Text S1

Figs. S1 to S27

Table S1 to S2

References

% \bibliographystyle{sciencemag}
% \bibliography{reference}% 
\putbib[reference]
\end{bibunit}

\noindent
\textbf{Acknowledgments: }We acknowledge Hua Tu for helpful discussions. 
\textbf{Funding: }This work is supported by Projects 12322501, 12575035, T2525011, 42450183, 12275020, 12135003, 12205025, and 42461144209 of the National Natural Science Foundation of China, and 2026NSFSCZY0124 of the Natural Science Foundation of Sichuan Province. J. F. acknowledges support from the Fundamental Research Funds for the Central Universities. 
\textbf{Author contributions: }Y.T. had the original idea for this work. Y.T. and J.F.F. designed and revised the workflow. H.T.X. and H.X.L. performed the study. H.T.X., H.X.L., J.F.F. and Y.T. contributed to the preparation of the manuscript.
\textbf{Competing interests: }The authors declare no competing interests.
\textbf{Data, code, and materials availability:}
The authors declare that the data supporting this study are available within the paper. A code implementation of the algorithm will be openly available upon acceptance of the manuscript.

\clearpage
\onecolumngrid
\setcounter{section}{0}
\setcounter{subsection}{0}
\setcounter{subsubsection}{0}
\setcounter{equation}{0}
\setcounter{figure}{0}
\setcounter{table}{0}

\renewcommand{\thesection}{S\arabic{section}}
\renewcommand{\thesubsection}{S\arabic{section}.\arabic{subsection}}
\renewcommand{\theequation}{S\arabic{equation}}
\renewcommand{\thefigure}{S\arabic{figure}}
\renewcommand{\thetable}{S\arabic{table}}

\RepeatTitleBlock

\vspace{1em}

\begin{bibunit}

\textbf{The PDF file includes:}

Supplementary Text S1

Figs. S1 to S27

Table S1 to S2

References

\newpage

\section{Details of Applying Large Deviation Theory to Extreme Precipitation}

In this section, we first describe the data preprocessing procedure in Subsection A, including the selection of representative grid points and the corresponding time series. Subsection B presents distribution fitting over different data ranges. Subsection C reports the return-time estimates obtained from large deviation theory and the enrichment procedure. Subsection D introduces the bias correction of future climate scenarios using quantile mapping. Subsection E summarizes the fitted parameter patterns. Subsection F analyzes precipitation entropy changes during ENSO transitions.

\subsection{Data Preprocessing}
\label{Data preprocessing}

We used daily precipitation data from the Beijing Climate Center Earth System Model version 1 (BCC-ESM1) model ~\cite{wu2020beijing}, which is part of the Coupled Model Intercomparison Project Phase 6 (CMIP6). The dataset covers the entire globe at a resolution of about 2.8125$^{\circ} \times$ 2.8125$^{\circ}$ and captures global precipitation patterns. Historical data from 1850 helped create a baseline. Our analysis included historical data and three future climate scenarios corresponding to the Shared Socioeconomic Pathways (SSP), SSP1-1.9, SSP4-3.4, and SSP5-8.5. These scenarios show different ways greenhouse gases could be emitted in the future. SSP1-1.9 represents a low-emissions pathway. It aligns with efforts to limit global warming to 1.5$^{\circ}$C. SSP4-3.4 reflects intermediate emissions. It has significant regional disparities in climate policy. SSP5-8.5 is a high-emission trajectory. It relies on fossil fuels and has minimal climate action. 

To more clearly evaluate the fitting performance of the distributions used in this paper for extreme precipitation, we selected 24 grid points worldwide corresponding to representative large cities on different continents, all of which have experienced extreme precipitation events, as shown in Fig.~S1. As the data source, we used daily precipitation output from CMIP6 for 1850–2015, as shown in Fig.~S2. For preprocessing, we considered two cases: using the raw data and removing days with daily precipitation below 0.1 mm. The latter was more suitable for most regions. For the extreme-precipitation range, we considered the upper tails above the 95th and 98th percentiles. This allowed us to study the tail of the full precipitation distribution while also providing a more comprehensive analysis of the tail above the 95th percentile.

\subsection{Fitting Distributions to Extreme Precipitation}
\label{Fitting of distributions}

In this section, we present the fitting performance of different distributions at 24 grid points under two preprocessing scenarios, as a supplement to Fig.~2. The paper uses Hellinger distance as the metric. Here, we also show the results using Jensen-Shannon divergence and Kullback-Leibler divergence in Fig. S3. First, we compare the fitting performance of distributions commonly used in previous studies with that of the distributions used in this paper for time series above the 95th percentile under two preprocessing conditions, retaining all precipitation days and removing days with precipitation below 0.1 mm. The results show the superior performance of the distributions used in this paper (Figs. S4, S5, S6, and S7). Next, we evaluate the fitting performance of the conventional distributions and the distributions used in this paper for data above the 95th and 98th percentiles after removing days with precipitation below 0.1 mm, as shown in Figs.~S6, S7, S8, and S9. The results show that fitting performance improves when days with precipitation below 0.1 mm are removed from the 95th percentile data and that the stable and Landau distributions outperform the alternatives in both comparisons.

\subsection{Return Time of Extreme Precipitation}
\label{Return time}

In this section, we present the full procedure and results for estimating return times using large deviation theory, as a supplement to Fig.~3. We first calculate the rate function for different block-averaging lengths, as shown in Fig.~S10, and find that the rate function curves gradually converge as the block length increases. Then we calculate the return time by selecting an appropriate block size. Finally, we enrich the return times obtained in Section IV by sampling from the fitted distributions in Section II, and we present global return-time results for different thresholds after enrichment under different preprocessing conditions and data ranges, thereby demonstrating the robustness of the method (Figs.~S11, S12, S13, S14, and S15). This section also extends the return time estimates for 24 representative grid points. The enriched return time curves preserve the goodness of fit to the historical data and substantially extend the prediction range, as shown in Figs.~S16, S17, and S18.

To validate the effectiveness of the fitted distribution, we assess precipitation return periods for 20 populous cities ~\cite{de2012vulnerability}, mainly in East Asia, South Asia, the Mediterranean Basin, and eastern North America, as a supplement to Table~I in the End Matter. Using the Landau distribution, predicted return times were compared with historical data, as shown in Table~S1. In general, the predictions for extreme precipitation levels of 100 mm remain highly accurate, although the error increases with longer return periods. In cities with high average precipitation, such as Manila and Mexico City, the frequency bias for a 30-year return period at 100 mm is minimal, with an error of only 7. This demonstrates the effectiveness of the Landau distribution in capturing and identifying extreme precipitation events that may occur in specific future periods.

Analyzing shifts in precipitation patterns over coming decades is integral to urban planning under climate change, as altered hydrological regimes may cascade through infrastructure networks and population centers. In this section, we examine the projected trends in precipitation, with a particular focus on the potential impacts of these changes on urban areas, especially those prone to extreme weather events. Based on Table S1, we have selected 10 key cities for further study. By examining three future climate scenarios, we aim to assist cities in preparing for and mitigating the impacts of changing precipitation patterns under climate uncertainty. Return periods are categorized as 5-, 10-, 20-, and 30-year periods. We then measure return frequencies for different precipitation magnitudes at different locations using three thresholds: 40 mm, 70 mm, and 100 mm.

Table S2 presents three rows for every city representing different climate scenarios. The first row is designated as SSP1-1.9, the second as SSP4-3.4, and the third as SSP5-8.5. Comparison across these cities shows that return frequencies decrease with the return period for all precipitation amounts. Under SSP5-8.5, return frequencies are notably higher for all precipitation amounts than under SSP1-1.9. This reflects the likelihood of an increase in the number of extreme precipitation events under extreme anthropogenic influences. Comparison with the results in Table~S1 shows that, for most cities, the predictions of our method align most closely with the return frequencies under SSP4-3.4. This scenario is characterized by lower mitigation challenges but higher adaptation challenges and is more reflective of the societal context in which we currently operate.

\subsection{Analysis of Future Data from Climate Models}
\label{Revision of future data}

In this section, we use quantile mapping to correct the data presented in Fig.~3 for future climate scenarios. We apply quantile mapping at representative grid points using precipitation data from the future scenarios SSP1-1.9, SSP4-3.4, and SSP5-8.5 together with historical CMIP6 precipitation data beginning in 1850, thereby obtaining correction factors for different precipitation amounts under the three scenarios. We then apply these three correction factors to the extended historical precipitation series obtained from distribution fitting to estimate return times over a wider precipitation range under the three future climate scenarios. We compare these estimates with the corresponding scenario statistics, which improves the reliability of the results (Figs.~S19, S20, S21, and S22).

\subsection{Fitting Parameters}
\label{Parameter laws}

To further show the underlying parameter patterns of the fitted distributions, we summarize the parameter behavior of the distributions used in this paper and find that the stability parameter $\alpha$ and the skewness parameter $\beta$ of the stable distribution are consistently close to 1 across the globe, as a supplement to Fig.~6 in the End Matter. Given this observation, we set both the stability parameter and the skewness parameter to 1, which causes the stable distribution to reduce to the Landau distribution. To determine how this reduction affects the results under different preprocessing settings, we conducted a more detailed analysis. We first compare the fitting performance of the Landau distribution with and without removing precipitation of less than 0.1 mm over the data range at the 95th percentile, as shown in Fig.~S23 (top and middle). We then evaluate how fitting the Landau distribution to data above the 95th and 98th percentiles affects performance over the corresponding quantile ranges, as shown in Fig.~S23 (middle and bottom). This section demonstrates the stability of the Landau distribution across different preprocessing conditions and data ranges.

\subsection{Precipitation Entropy Changes during ENSO}
\label{Precipitation entropy changes during ENSO}

The fluctuation theorem (FT), a fundamental principle in non-equilibrium statistical mechanics, establishes a relationship between the probabilities of positive and negative entropy production. We apply the FT to analyze precipitation entropy changes during El Ni\~{n}o-Southern Oscillation (ENSO) transitions from 1850 to 2015, aiming to determine whether the observed entropy production rates adhere to the theoretical symmetry predicted by the FT. The cumulative distribution functions (CDFs) of entropy production rates are computed for both full trajectory and extremes-only periods, providing a framework to assess the symmetry about $ \Delta S = 0 $ (Figs. S24, S25, S26 and S27).

Global precipitation averages are calculated by spatially averaging precipitation over all grid points for each time step. Let $ \text{pr}_{i,j}(t) $ represent the precipitation rate at grid point $(i,j)$ at time $ t $, and let $ N $ denote the total number of grid points. The global average precipitation at time $ t $ is computed as

\begin{equation}
\text{pr}_{\text{avg}}(t) = \frac{1}{N} \sum_{i,j} \text{pr}_{i,j}(t).
\end{equation}

To ensure comparability across different time periods, the global average precipitation is normalized by subtracting its mean and dividing by its standard deviation,

\begin{equation}
\text{pr}_{\text{norm}}(t) = \frac{\text{pr}_{\text{avg}}(t) - \mu}{\sigma},
\end{equation}
where $ \mu $ and $ \sigma $ are the mean and standard deviation of $ \text{pr}_{\text{avg}}(t) $, respectively. 

Entropy is calculated based on the temporal variability of the normalized precipitation data. The precipitation time series is treated as a discrete stochastic process, and entropy production is quantified using the fluctuation of precipitation values over time. For a given time interval $ \Delta t $, the change in precipitation, $ \Delta \text{pr}_{\text{norm}}(t) $, is defined as

\begin{equation}
\Delta \text{pr}_{\text{norm}}(t) = \text{pr}_{\text{norm}}(t + \Delta t) - \text{pr}_{\text{norm}}(t).
\end{equation}

To calculate entropy production, the probability distribution function (PDF) of $ \Delta \text{pr}_{\text{norm}}(t) $ is estimated. Let $ P(\Delta \text{pr}) $ denote the probability of a precipitation change $ \Delta \text{pr} $. Following principles of non-equilibrium statistical mechanics, the entropy production $ S $ associated with a precipitation change is given by

\begin{equation}
S = -\sum_{\Delta \text{pr}} P(\Delta \text{pr}) \ln P(\Delta \text{pr}),
\end{equation}
where the summation is taken over all possible precipitation changes. This entropy quantifies the uncertainty or disorder in the precipitation fluctuations.

To evaluate compliance with the FT, the symmetry of the entropy production is analyzed. The FT predicts that the ratio of probabilities of positive and negative entropy production satisfies

\begin{equation}
\frac{P(\Delta S)}{P(-\Delta S)} = e^{\Delta S/k_B},
\end{equation}
where $ k_B $ is the Boltzmann constant. The CDFs of entropy production are computed to observe the symmetry about $ \Delta S = 0 $. If the FT holds, the CDFs should show balanced distributions of positive and negative entropy production.

The CDFs derived from the full trajectory periods show that entropy production during ENSO transitions exhibits no significant asymmetry around $ \Delta S = 0 $. The distributions are continuous and symmetric in most periods, indicating that the probabilities of positive and negative entropy production rates are comparable. The smoothness of the CDFs further suggests that entropy production follows a consistent pattern over the full trajectory of the ENSO periods, with no evident dominance of positive or negative values. The extremes-only analysis emphasizes larger fluctuations in entropy production, yet it does not show significant deviations from the fluctuation theorem. The step-like nature of the CDFs in the extremes-only periods reflects the sparsity of extreme entropy values, but the symmetry about $ \Delta S = 0 $ remains preserved. The probability of observing large positive and negative entropy production events appears balanced, suggesting that the FT may hold even under extreme conditions during ENSO events. The lack of clear asymmetry implies that the system's dynamics during extreme events do not strongly violate the FT's predictions.

The results from both full trajectory and extremes-only periods suggest that the precipitation entropy production during ENSO transitions is consistent with the fluctuation theorem. The CDFs show no significant asymmetry around $ \Delta S = 0 $, indicating that the probabilities of positive and negative entropy production are comparable. These findings imply that the ENSO system, while inherently non-equilibrium, may exhibit statistical properties that align with the FT. 

\clearpage
\section*{Supplementary Figures}

\begin{figure}[H]
    \centering
    \includegraphics[width=0.5\linewidth]{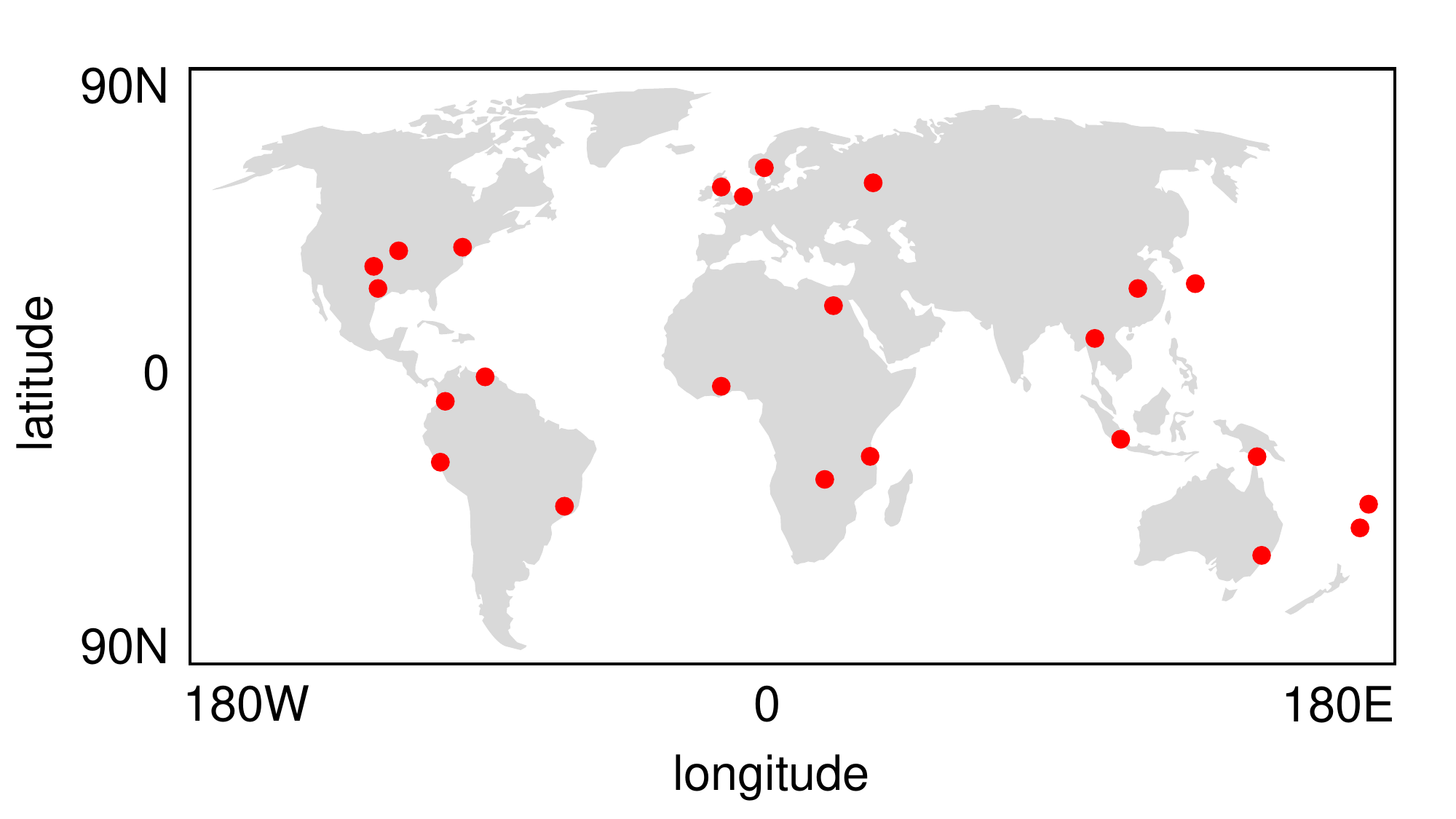}
    \caption{\textbf{Locations of the selected grid points (red), as a supplement to Fig.~1.} The 24 selected grid points correspond to representative large cities on different continents that have experienced extreme precipitation events and are used to illustrate the fit of the extreme precipitation distributions.}
\end{figure}

\clearpage
\begin{figure}[H]
    \centering
    \includegraphics[width=\linewidth]{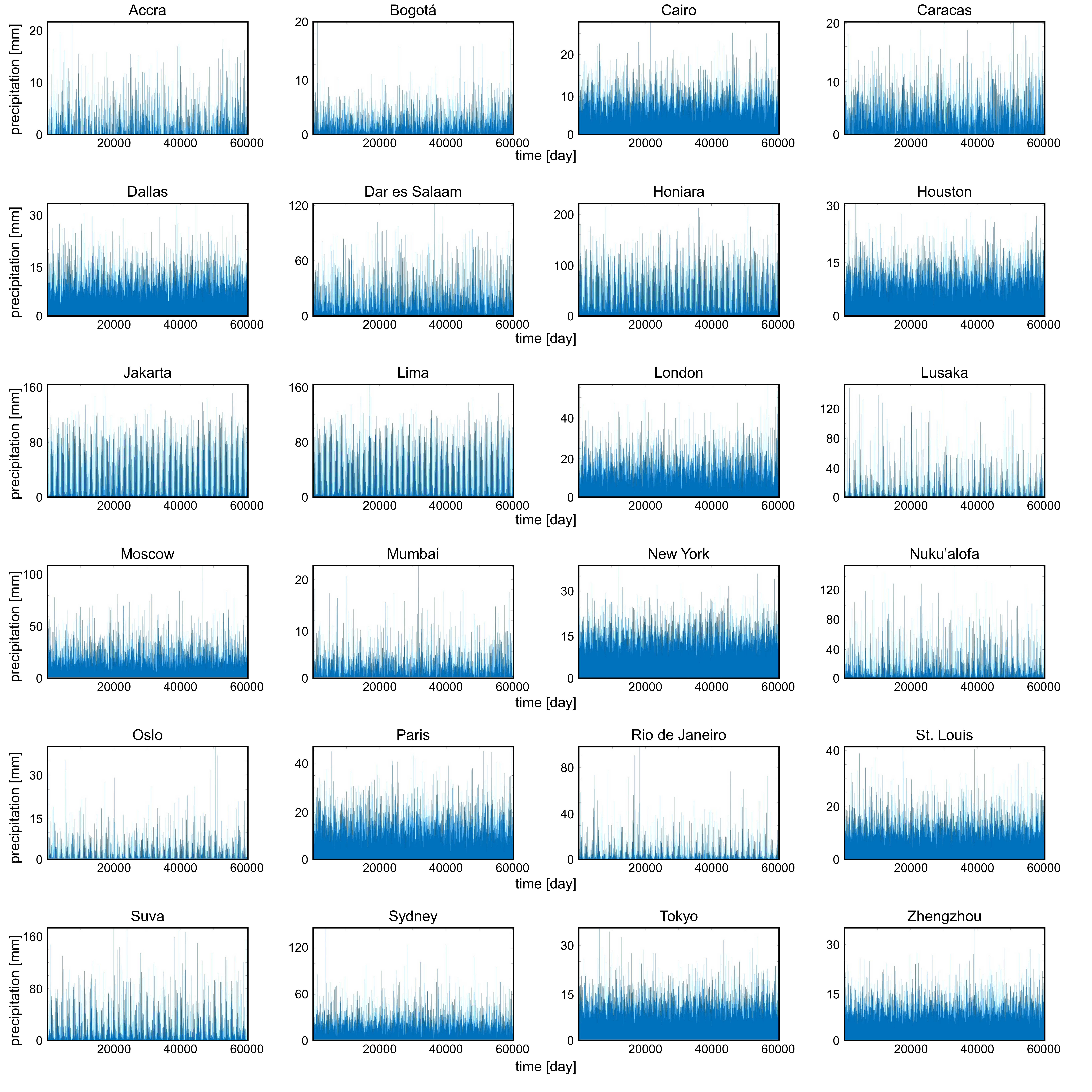}
    \caption{\textbf{Precipitation time series for the grid points over the data interval, as a supplement to Fig. ~1.} The dataset comprises daily precipitation time series spanning 60,225 days from 1850 to 2015 from Coupled Model Intercomparison Project Phase (CMIP6). The daily precipitation time series at these grid points show that almost every location experiences occasional extreme events that greatly exceed the long term average observed on most days. The frequency of extreme precipitation events also varies substantially across grid points.}
\end{figure}

\clearpage
\begin{figure}[H]
    \centering
    \includegraphics[width=\linewidth]{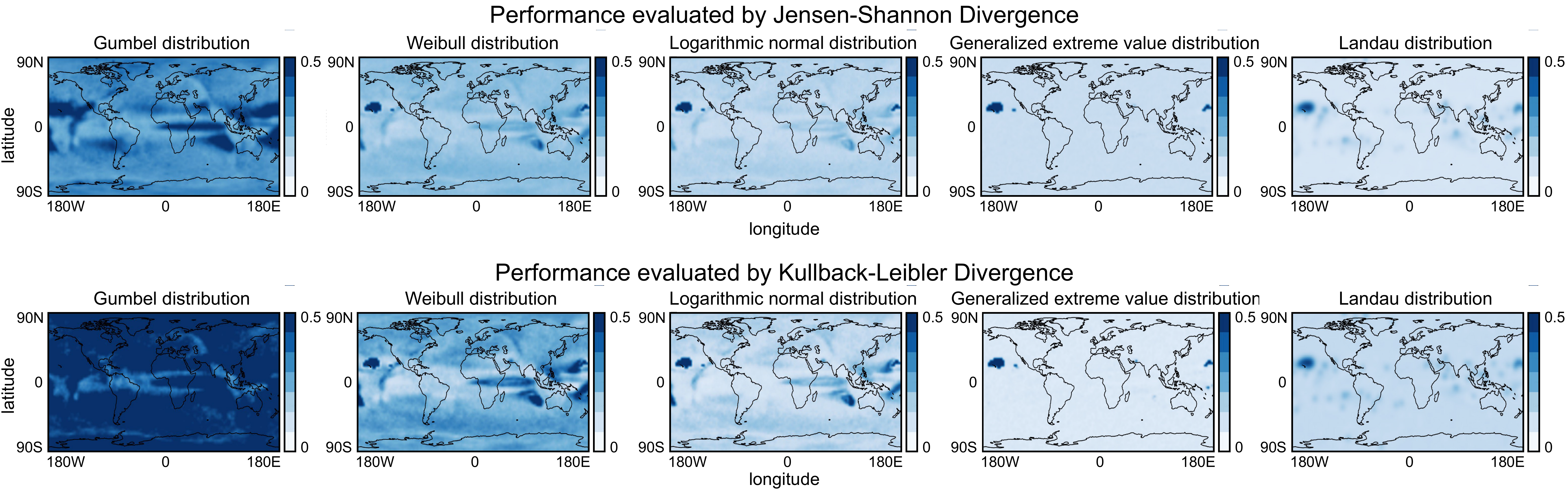}
    \caption{\textbf{Performance of distributions in modeling extreme precipitation events, as a supplement to Fig. ~2.} The Gumbel, Weibull, Logarithmic normal, generalized extreme value and Landau distributions are used for modeling 95\% precipitation events and evaluated by Jensen-Shannon and Kullback-Leibler Divergence.}
\end{figure}

\begin{figure}[H]
    \centering
    \includegraphics[width=\linewidth]{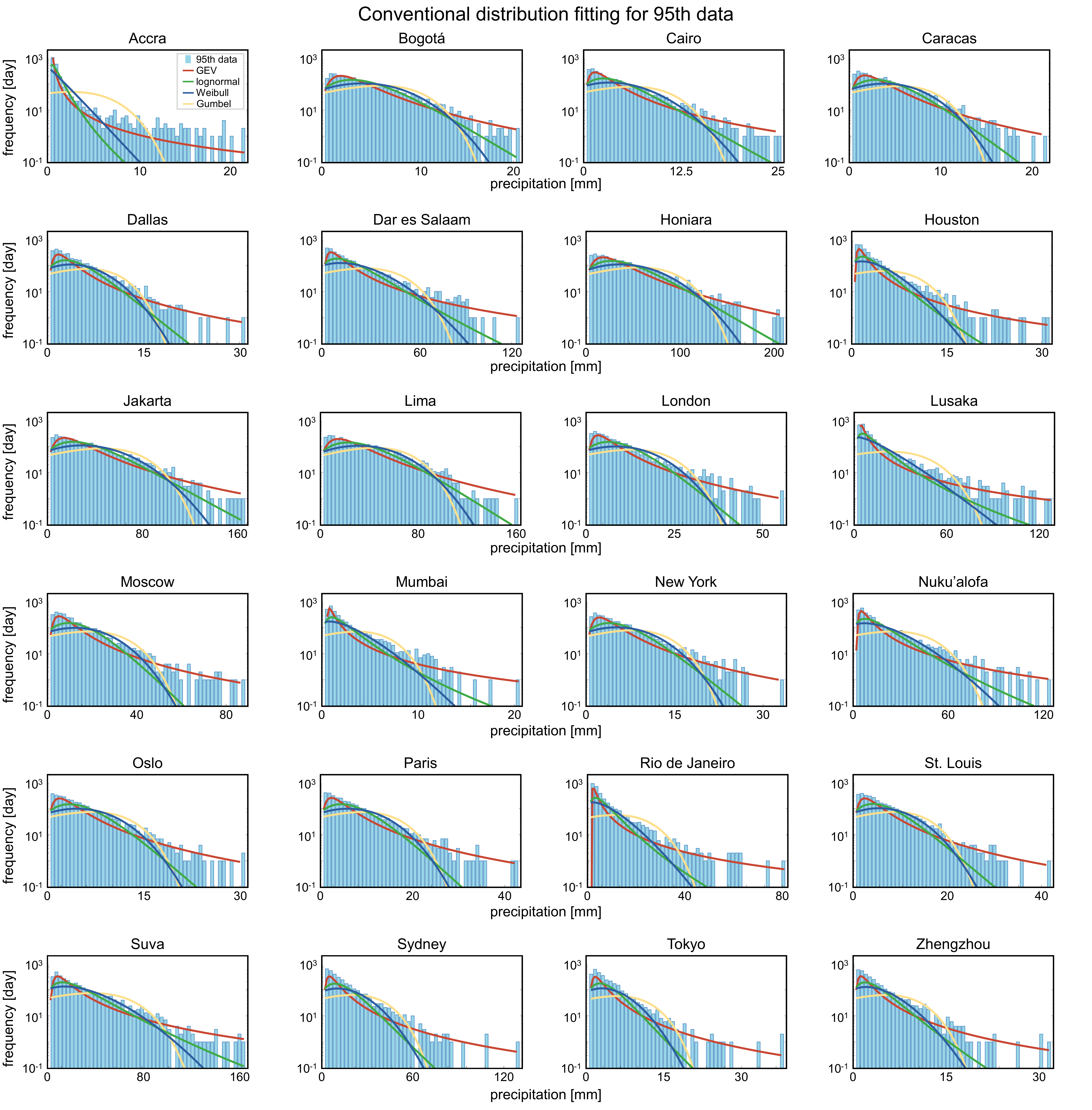}
    %Fitted conventional distribution of the grid points. 
    \caption{\textbf{95th percentile conventional distribution fitted for 24 grid points, as a supplement to Fig. ~2.} The Gumbel, Weibull, logarithmic normal (lognormal), Generalized Extreme Value (GEV), and Landau distributions are fitted to precipitation data above the 95th percentile and evaluated using the Jensen-Shannon and Kullback-Leibler divergences.}
\end{figure}

\begin{figure}[H]
    \centering
    \includegraphics[width=\linewidth]{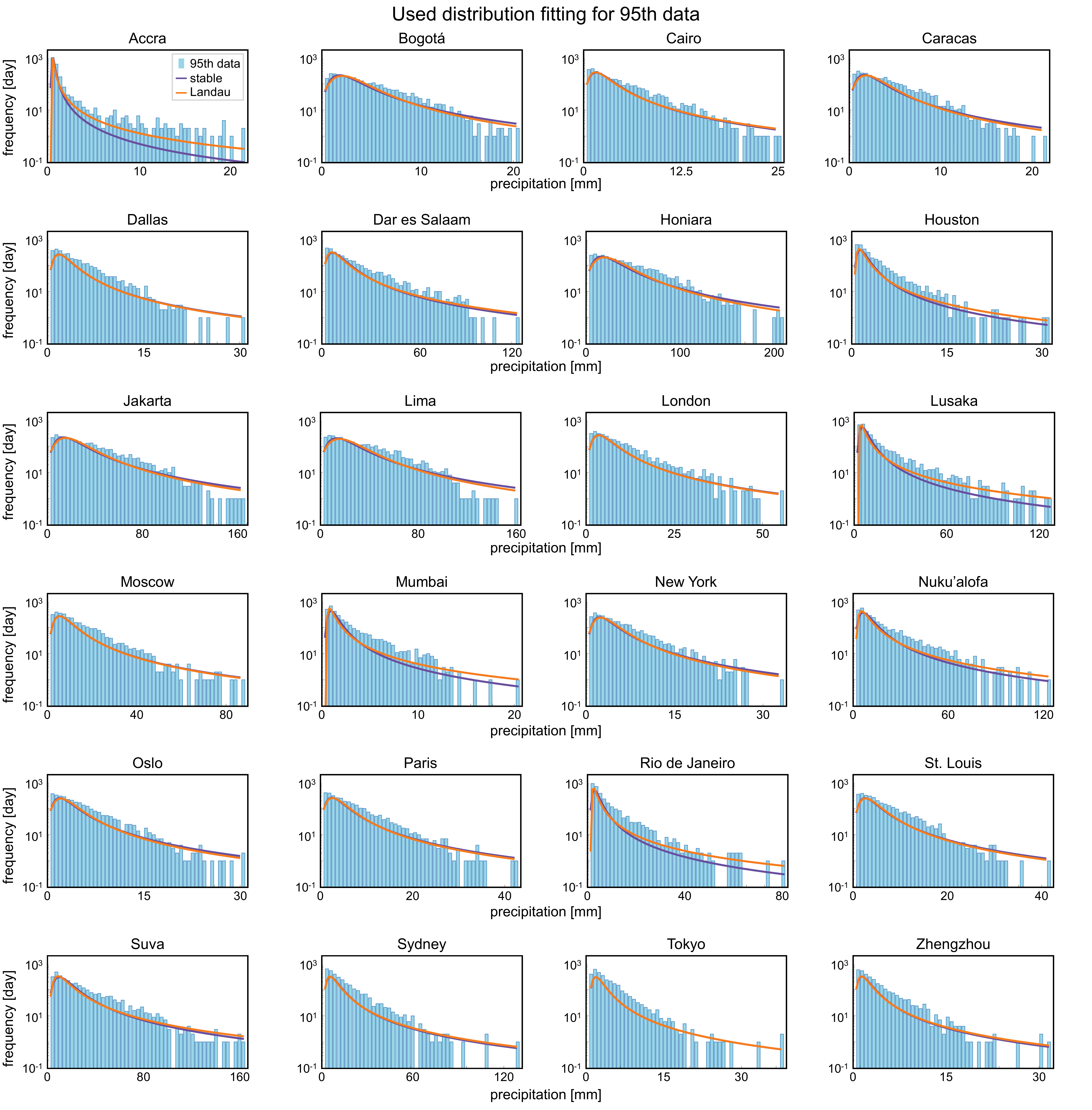}
    %Fitted used distribution of the grid points
    \caption{\textbf{95th percentile used distribution fitted for 24 grid points, as a supplement to Fig. ~2.} The distributions for stable (red) and Landau (green) distribution are shown.}
\end{figure}

\begin{figure}[H]
    \centering
    \includegraphics[width=\linewidth]{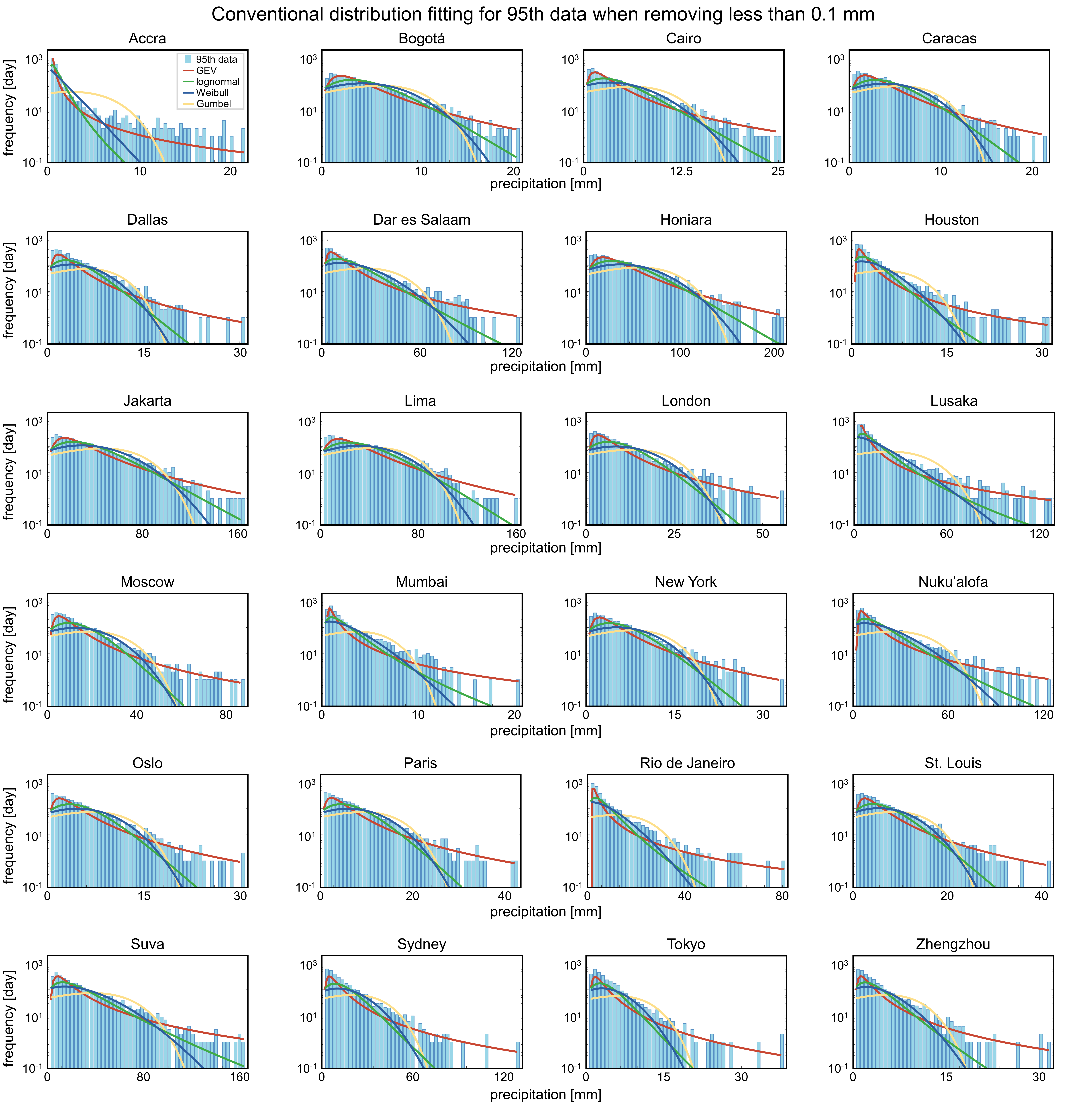}
    \caption{\textbf{95th percentile conventional distribution fitted for 24 grid points when removing less than 0.1 mm of precipitation days, as a supplement to Fig. ~2.} The fitted GEV (red), lognormal (dark green), Weibull (blue), Gumbel (yellow), and Generalized Pareto (GP, light green) distributions are shown.}
\end{figure}

\begin{figure}[H]
    \centering
    \includegraphics[width=\linewidth]{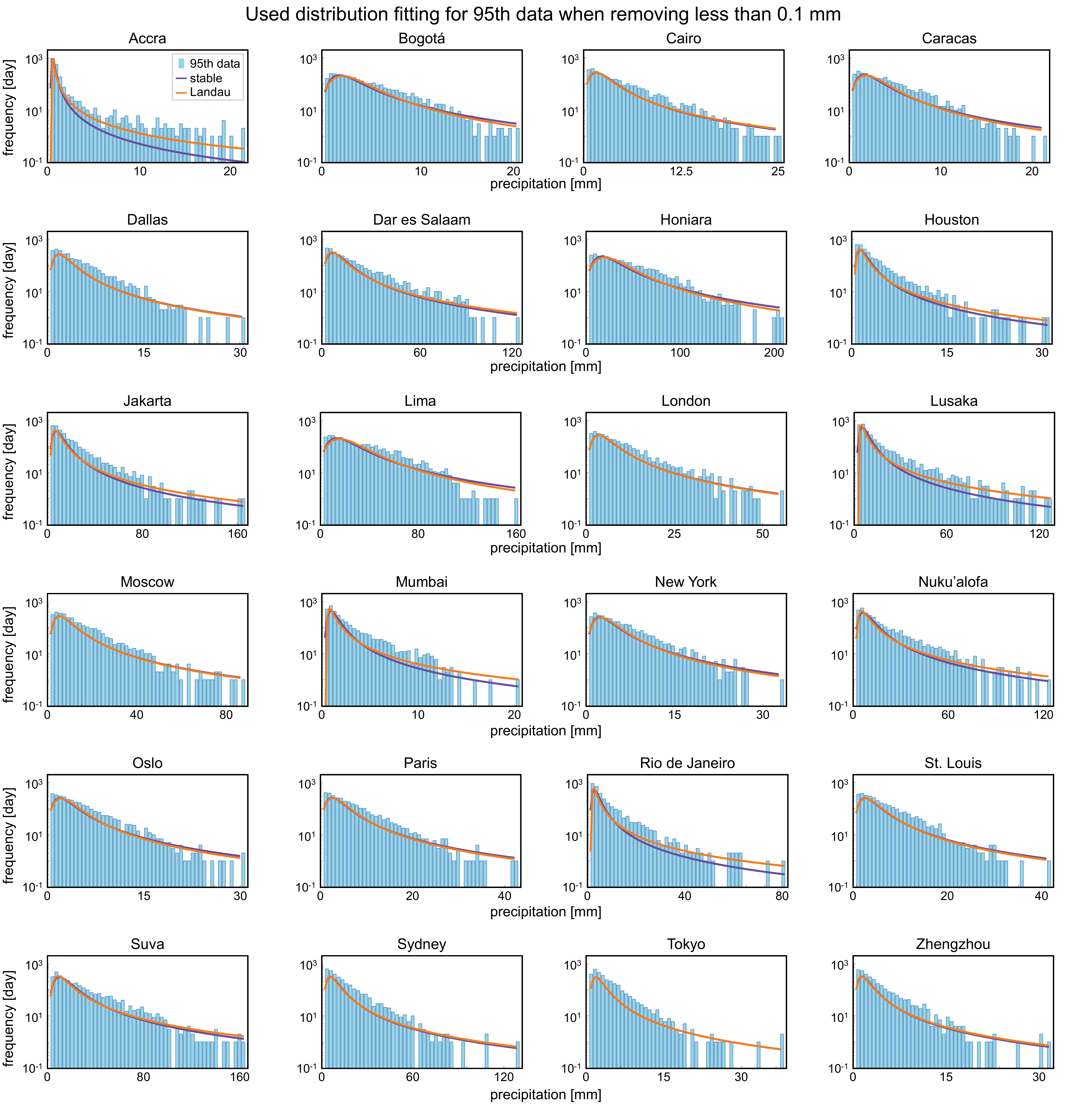}
    %Fitted used distribution of the grid points
    \caption{\textbf{95th percentile used distribution fitted for 24 grid points when removing less than 0.1 mm of precipitation days, as a supplement to Fig. ~2.} The distributions for stable (red) and Landau (green) distribution are shown.}
\end{figure}

\begin{figure}[H]
    \centering
    \includegraphics[width=\linewidth]{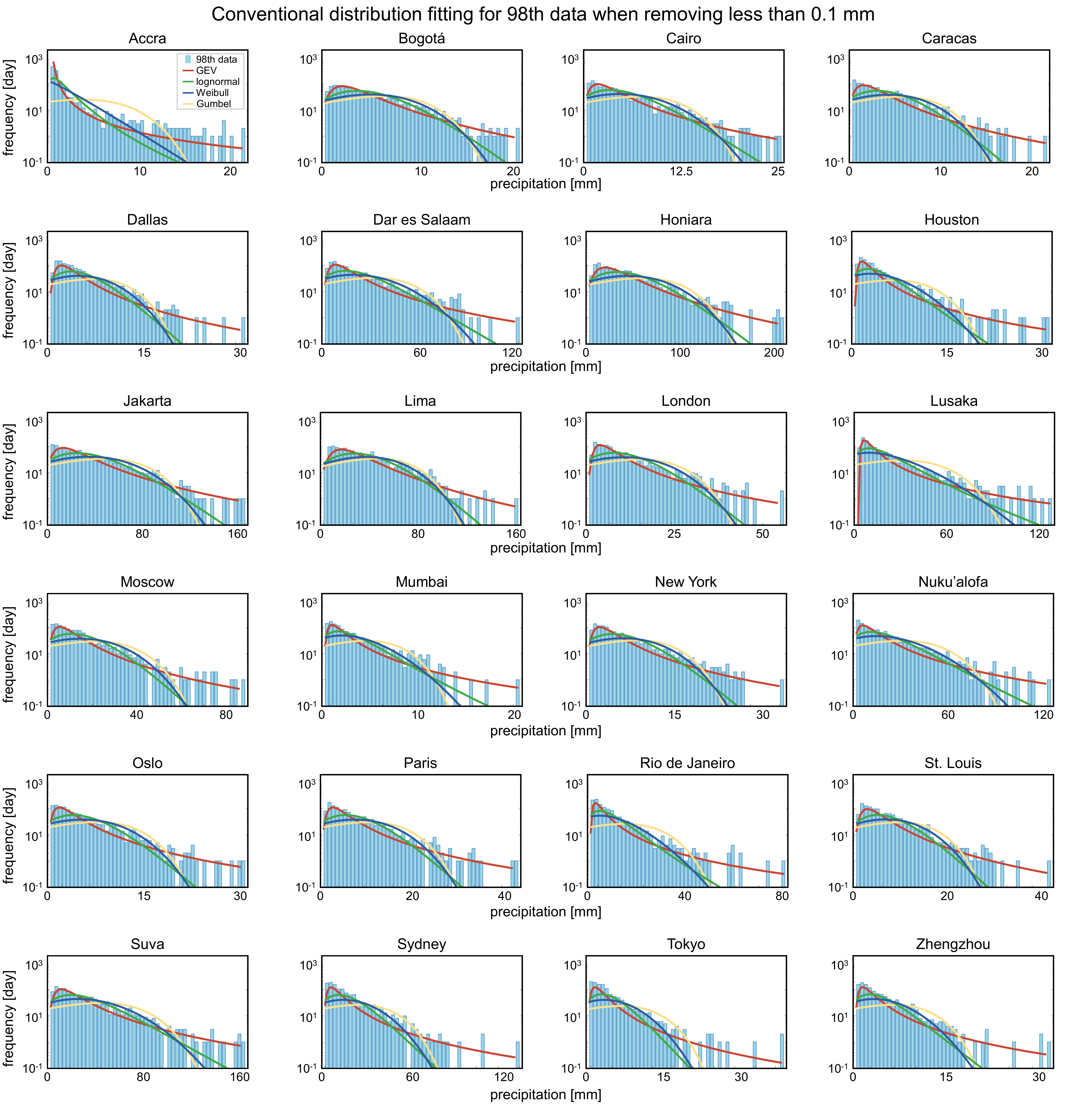}
    %Fitted conventional distribution of the grid points.
    \caption{\textbf{98th percentile conventional distribution fitted for 24 grid points when removing less than 0.1 mm of precipitation days, as a supplement to Fig. ~2.} The fitted GEV (red), lognormal (dark green), Weibull (blue), Gumbel (yellow), and GP (light green) distributions are shown.}
\end{figure}

\begin{figure}[H]
    \centering
    \includegraphics[width=\linewidth]{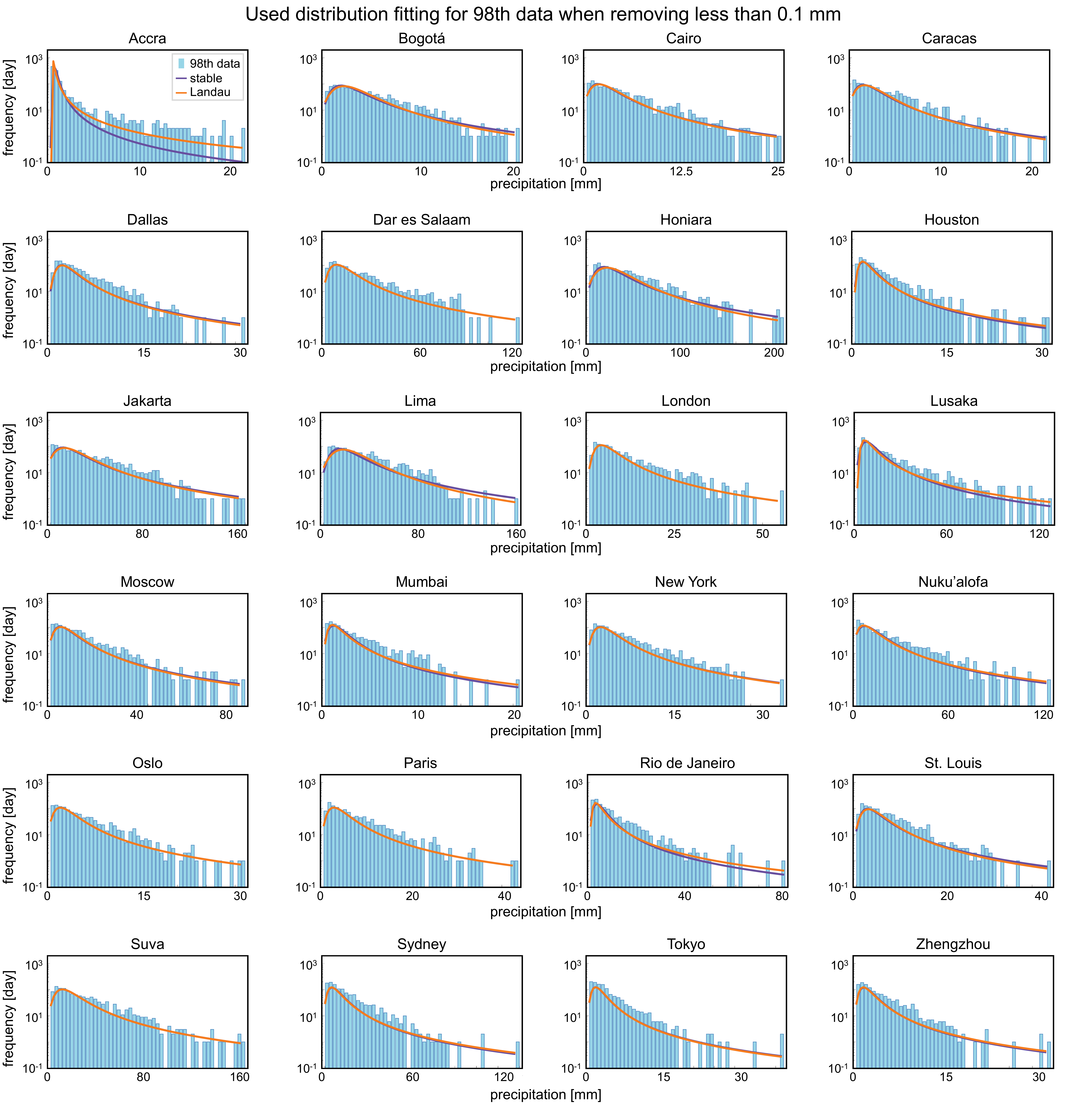}
    %Fitted used distribution of the grid points
    \caption{\textbf{98th percentile used distribution fitted for 24 grid points when removing less than 0.1 mm of precipitation days, as a supplement to Fig. ~2.} The distributions for stable (red) and Landau (green) distribution are shown.}
\end{figure}

\begin{figure}[H]
    \centering
    \includegraphics[width=\linewidth]{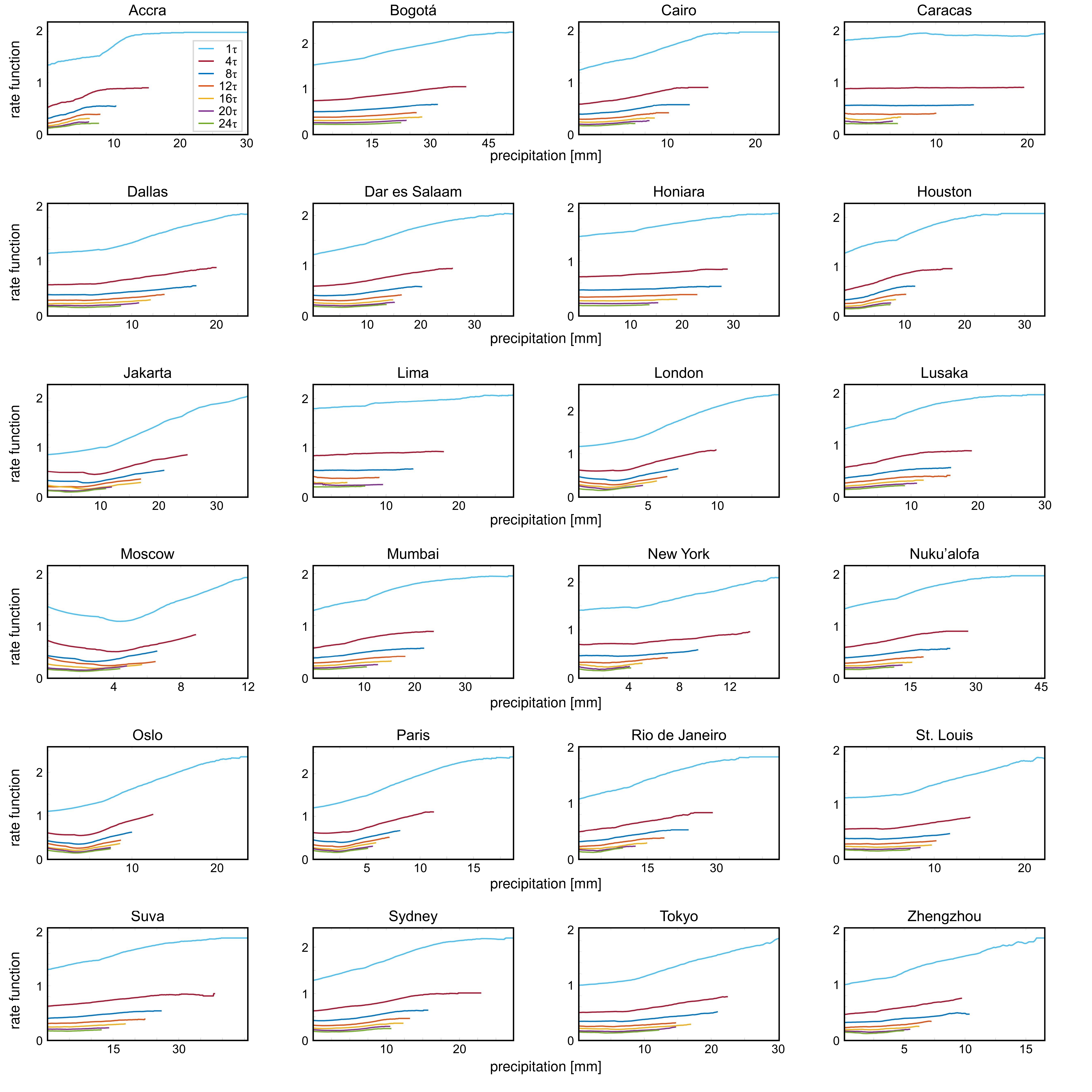}
    \caption{\textbf{Rate functions at selected grid points at different block average lengths, as a supplement to Fig. ~3.} 
    We compute $I(n)$ for block-averaging lengths $n = 1\tau, 4\tau, ..., 24\tau$ under the condition that all precipitation days are retained. As the average length n increases, the rate function curve gradually converges.}
\end{figure}

\begin{figure}[H]
    \centering
    \includegraphics[width=\linewidth]{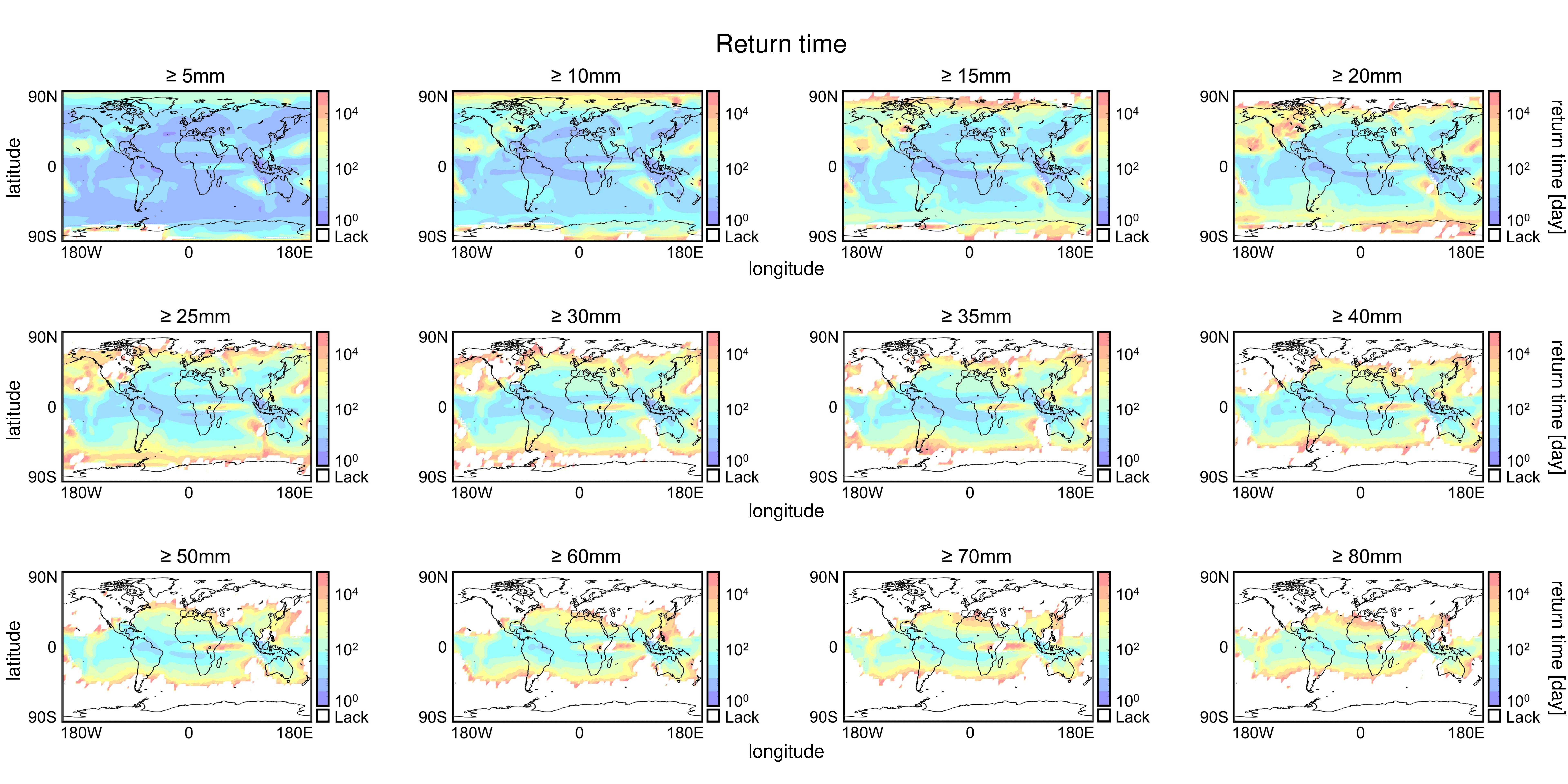}
    \caption{\textbf{Spatial distributions of return time for $\tau = 1$ with all precipitation days retained, as a supplement to Fig.~3.} Thresholds from $\ge 5$ to $\ge 80$ mm are shown at 5-mm intervals, except that thresholds from $\ge 50$ to $\ge 80$ mm are shown at 10-mm intervals. The number of days required for precipitation to return to a specific threshold generally increases with increasing latitude. However, more missing values appear at the North and South Poles when the precipitation threshold is high. Furthermore, it can be seen that the increase in return time is slower for the oceans than for the continents.}
\end{figure}

\begin{figure}[H]
    \centering
    \includegraphics[width=\linewidth]{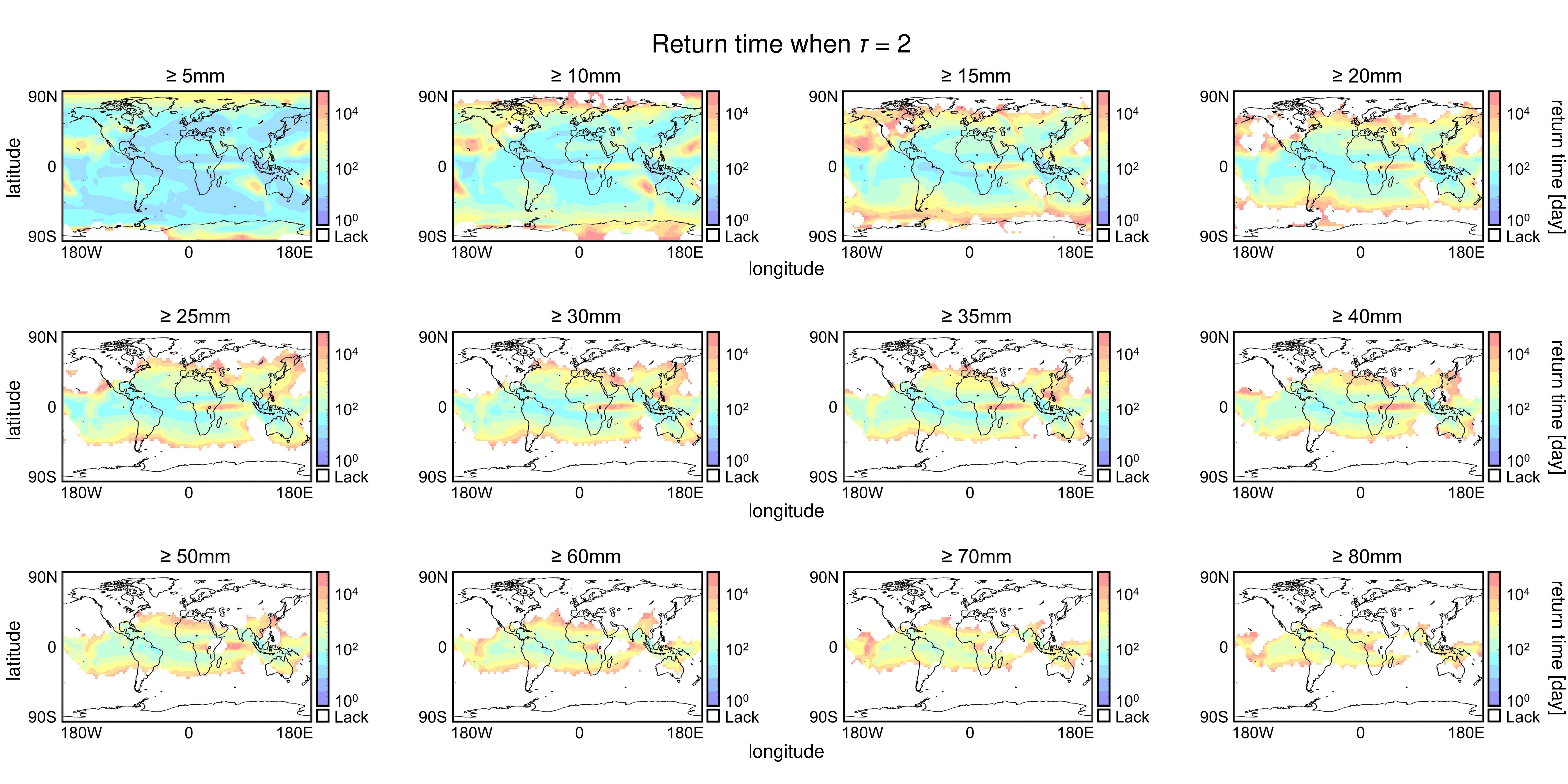}
    \caption{\textbf{Spatial distributions of return time for $\tau = 2$ with all precipitation days retained, as a supplement to Fig.~3.} Thresholds from $\ge 5$ to $\ge 80$ mm are shown in the figure at 5 mm intervals, where the $\ge 50$ to $\ge 80$ mm part threshold is 10 mm. The spatial distributions are similar to those shown in Fig.~S11. However, because of the error introduced by the block averaging length, the spatial distribution contains more missing values than in the case $\tau = 1$, and the return times are also longer.}
\end{figure}

\begin{figure}[H]
    \centering
    \includegraphics[width=\linewidth]{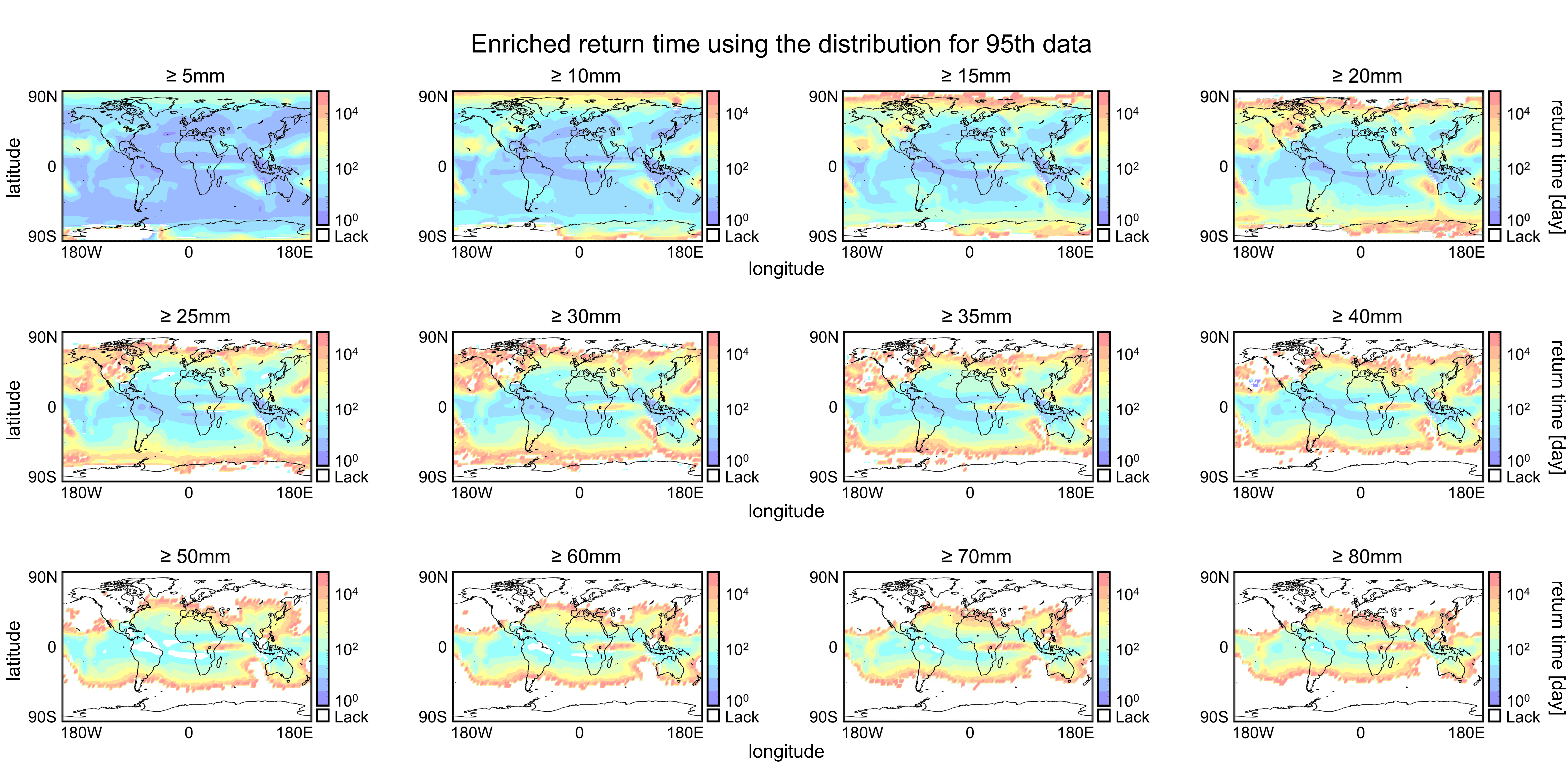}
    \caption{\textbf{Spatial distributions of return time after sampling from fits to 95th percentile distributions with all precipitation days retained, as a supplement to Fig.~3.} Thresholds from $\ge 5$ to $\ge 80$ mm are shown in the figure at 5 mm intervals, where the $\ge 50$ to $\ge 80$ mm part threshold is 10 mm. Compared with Fig.~S12, return times increase markedly after enrichment. This is particularly evident at the poles and in other low-precipitation regions, where missing values occur at small thresholds. At larger thresholds, the enriched return time widens slightly towards the poles, in comparison to when there is no enrichment.}
\end{figure}

\begin{figure}[H]
    \centering
    \includegraphics[width=\linewidth]{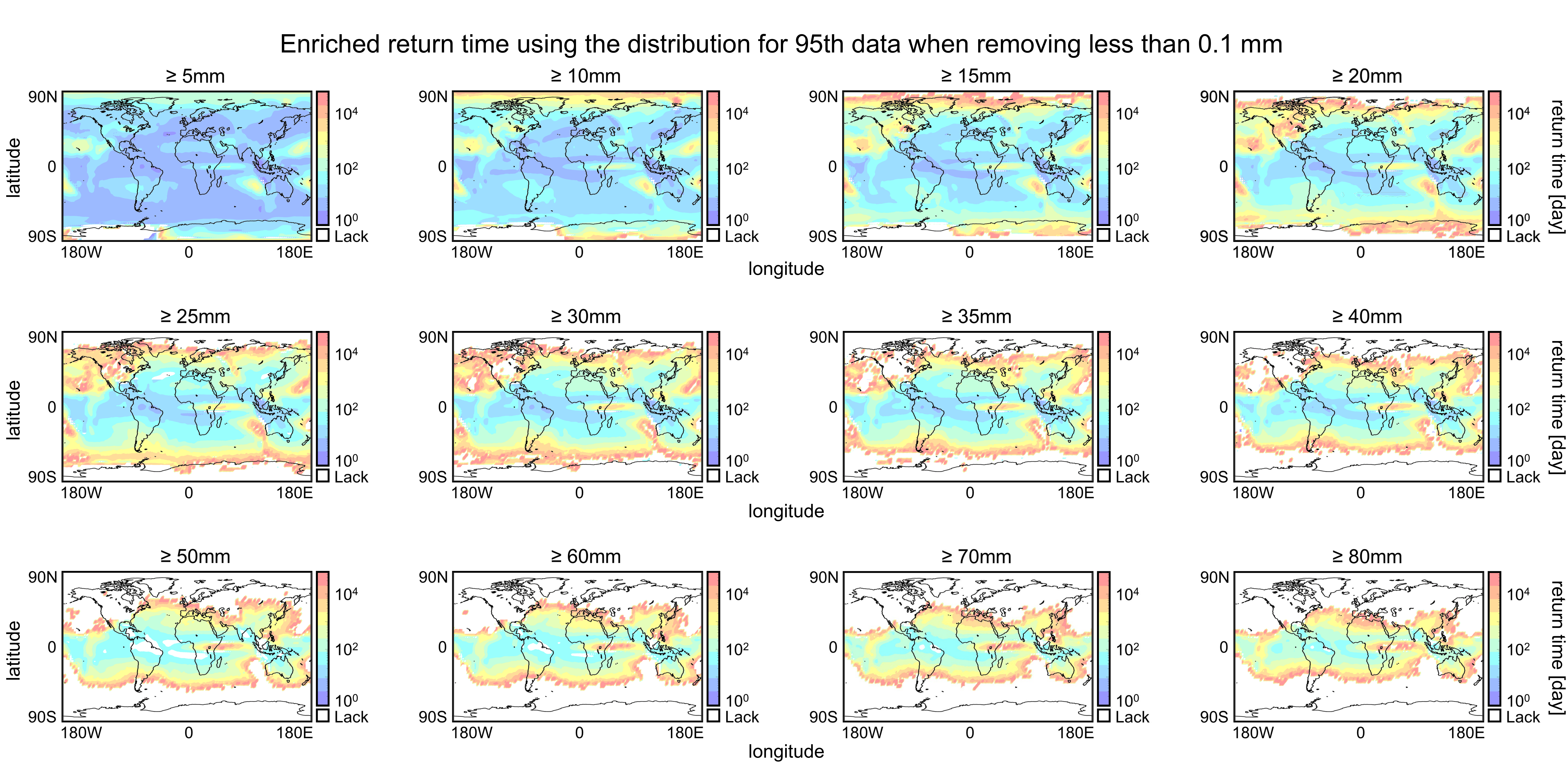}
    \caption{\textbf {Spatial distributions of return time after sampling from fits to 95th percentile distributions after removing days with precipitation below 0.1 mm, as a supplement to Fig.~3.} Thresholds from $\ge 5$ to $\ge 80$ mm are shown in the figure at 5 mm intervals, where the $\ge 50$ to $\ge 80$ mm part threshold is 10 mm. Compared with Fig.~S13, the enriched return times remain relatively consistent despite the change in data range. This suggests that enrichment is a stable process, and that the use of mainstream data preprocessing methods does not significantly impact the results.}
\end{figure}

\begin{figure}[H]
    \centering
    \includegraphics[width=\linewidth]{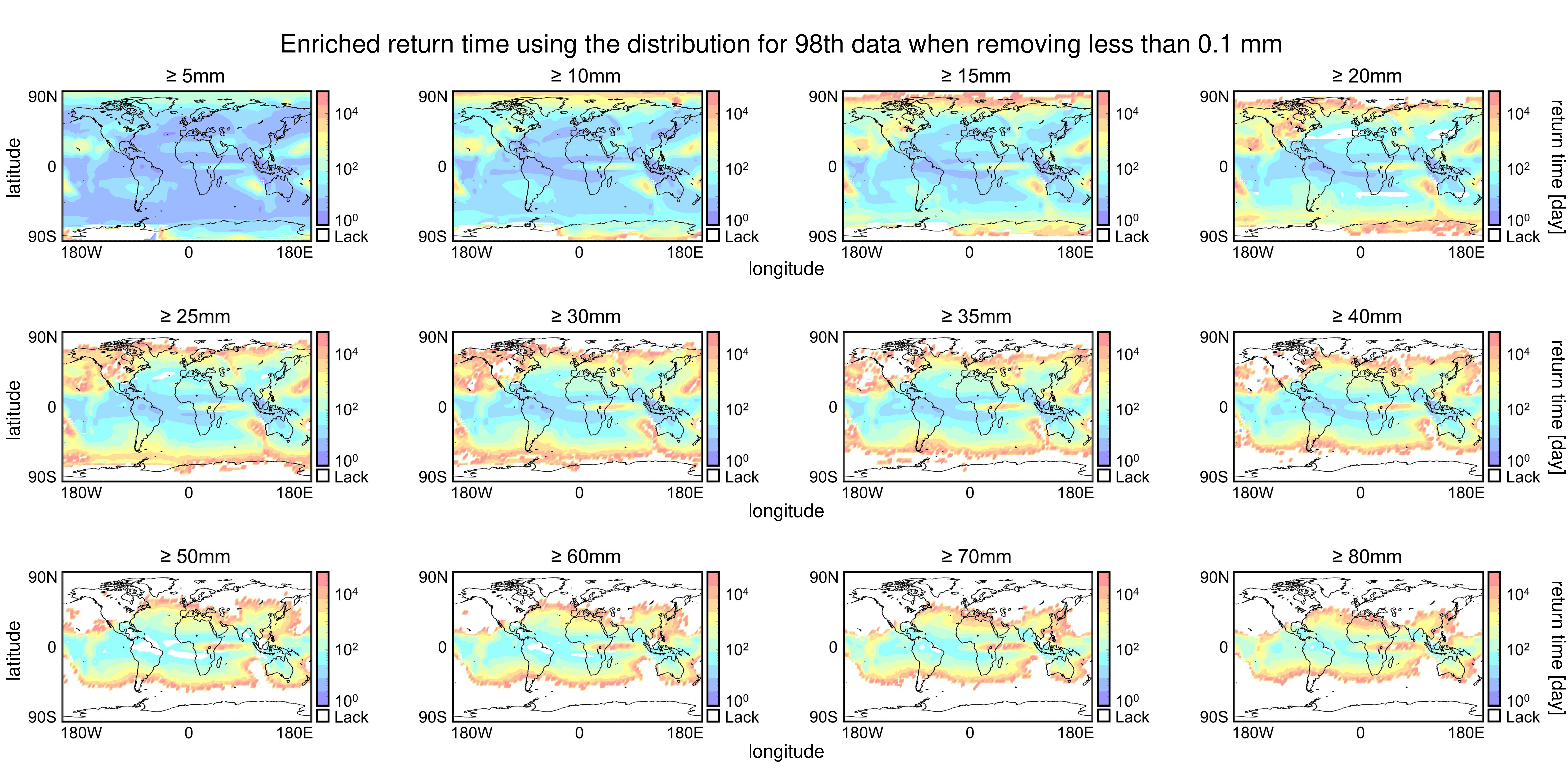}
    \caption{\textbf {Spatial distributions of return time after sampling from fits to 98th percentile distributions after removing days with precipitation below 0.1 mm, as a supplement to Fig.~3.} Thresholds from $\ge 5$ to $\ge 80$ mm are shown in the figure at 5 mm intervals, where the $\ge 50$ to $\ge 80$ mm part threshold is 10 mm. Comparison of the relevant panels shows that when only data above the 98th percentile are fitted, sampled, and enriched, the effect at the poles is less pronounced than when data above the 95th percentile are used.}
\end{figure}

\begin{figure}[H]
    \centering
    \includegraphics[width=\linewidth]{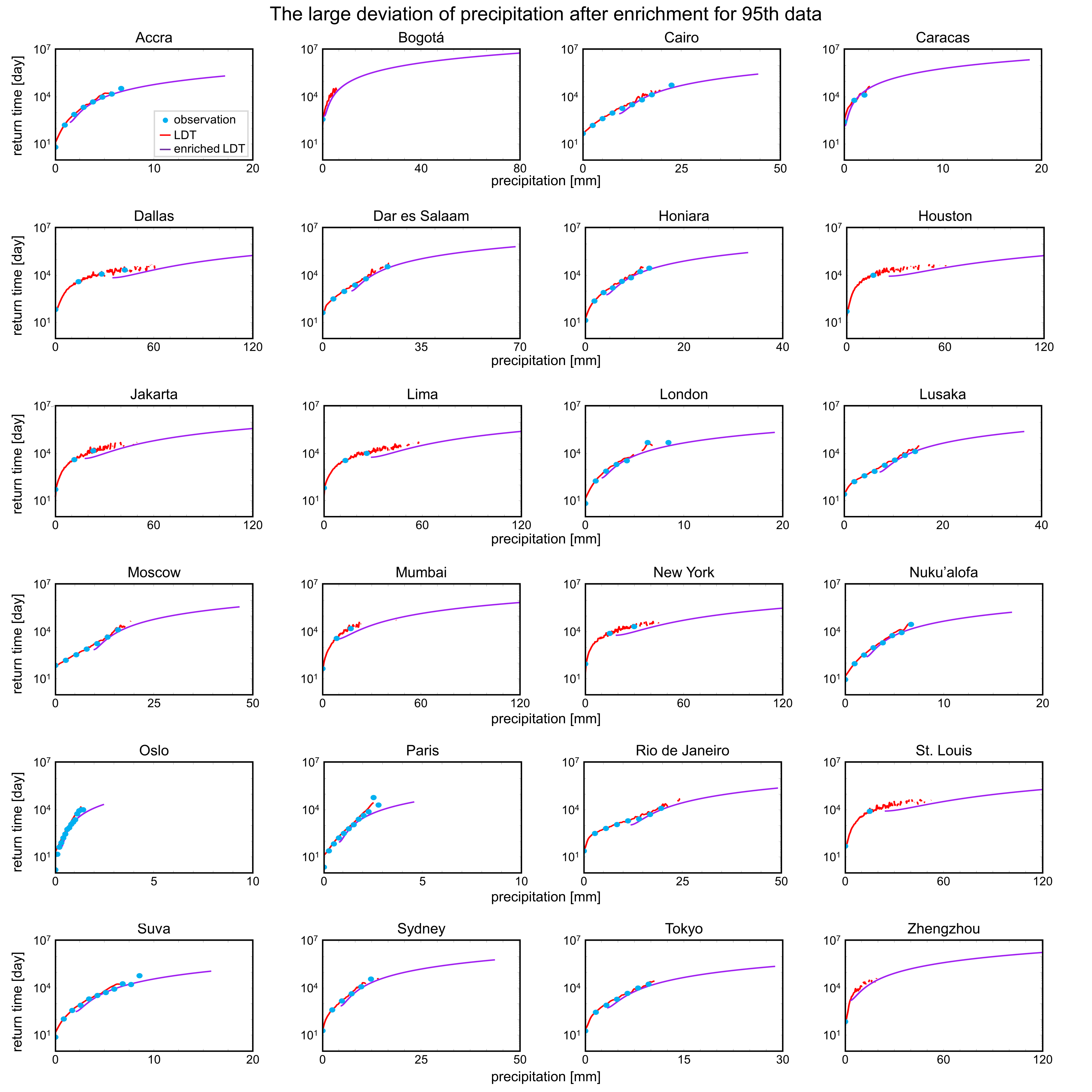}
    \caption{\textbf{Calculation of the large deviation of precipitation after enrichment by sampling from fits to 95th percentile distributions, as a supplement to Fig. ~3.} Historical precipitation data (blue points), return times calculated using large-deviation theory (LDT, red line), and enriched return times obtained by sampling (purple line) are shown for selected grid points, thereby extending the prediction range at each location. To reduce errors, enriched return times to the left of the distribution peak were removed.}
\end{figure}

\begin{figure}[H]
    \centering
    \includegraphics[width=\linewidth]{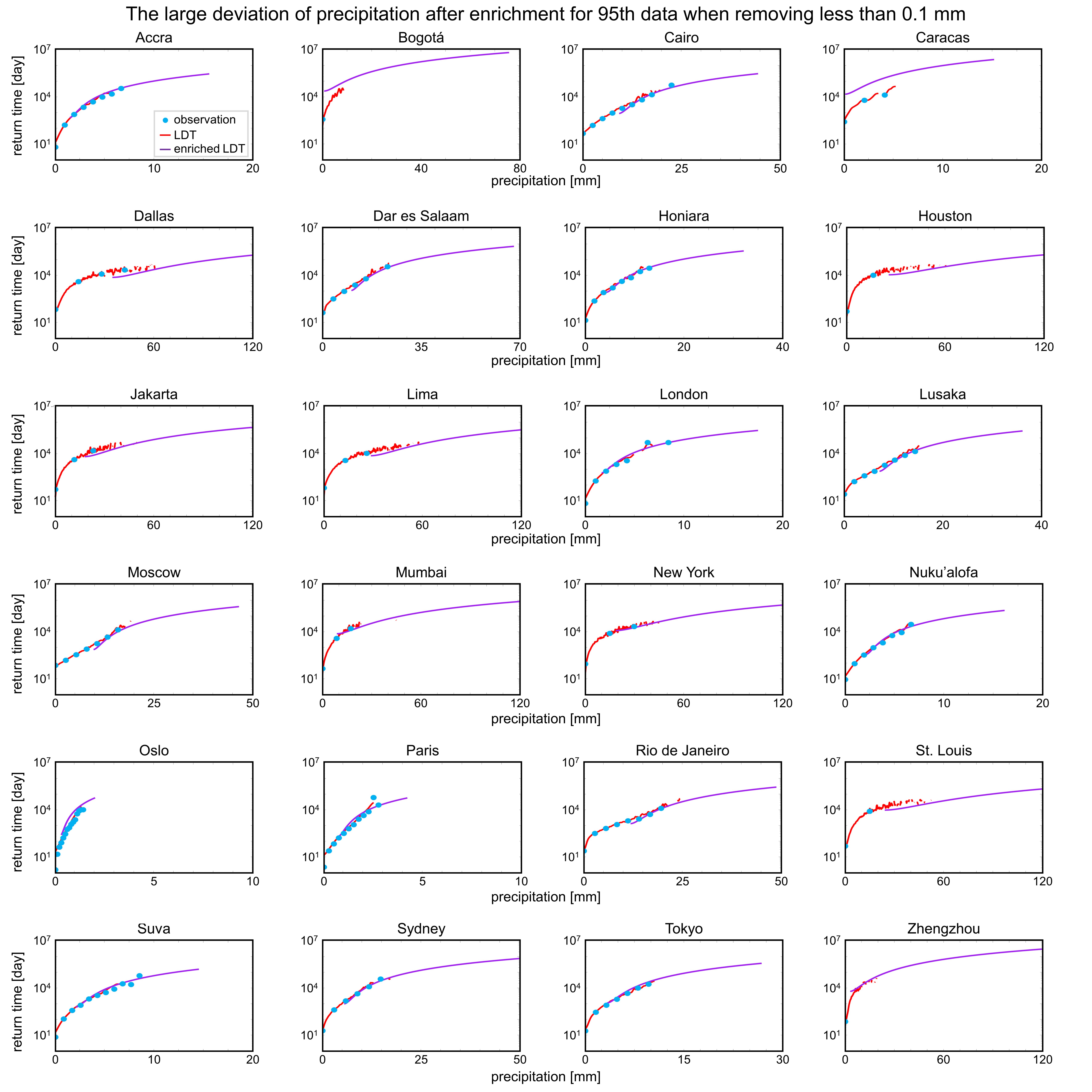}
    \caption{\textbf{Calculation of the large deviation of precipitation after enrichment by sampling from fits to 95th percentile distributions after removing less than 0.1 mm of precipitation days, as a supplement to Fig. ~3.} Historical precipitation data (blue points) and return time calculated using large deviation theory (LDT, red line) and enriched (purple line) by sampling at selected grid points, expanding the return time prediction range of the location. To reduce errors, the enriched return time were removed from the left side of the distribution peak.}
\end{figure}

\begin{figure}[H]
    \centering
    \includegraphics[width=\linewidth]{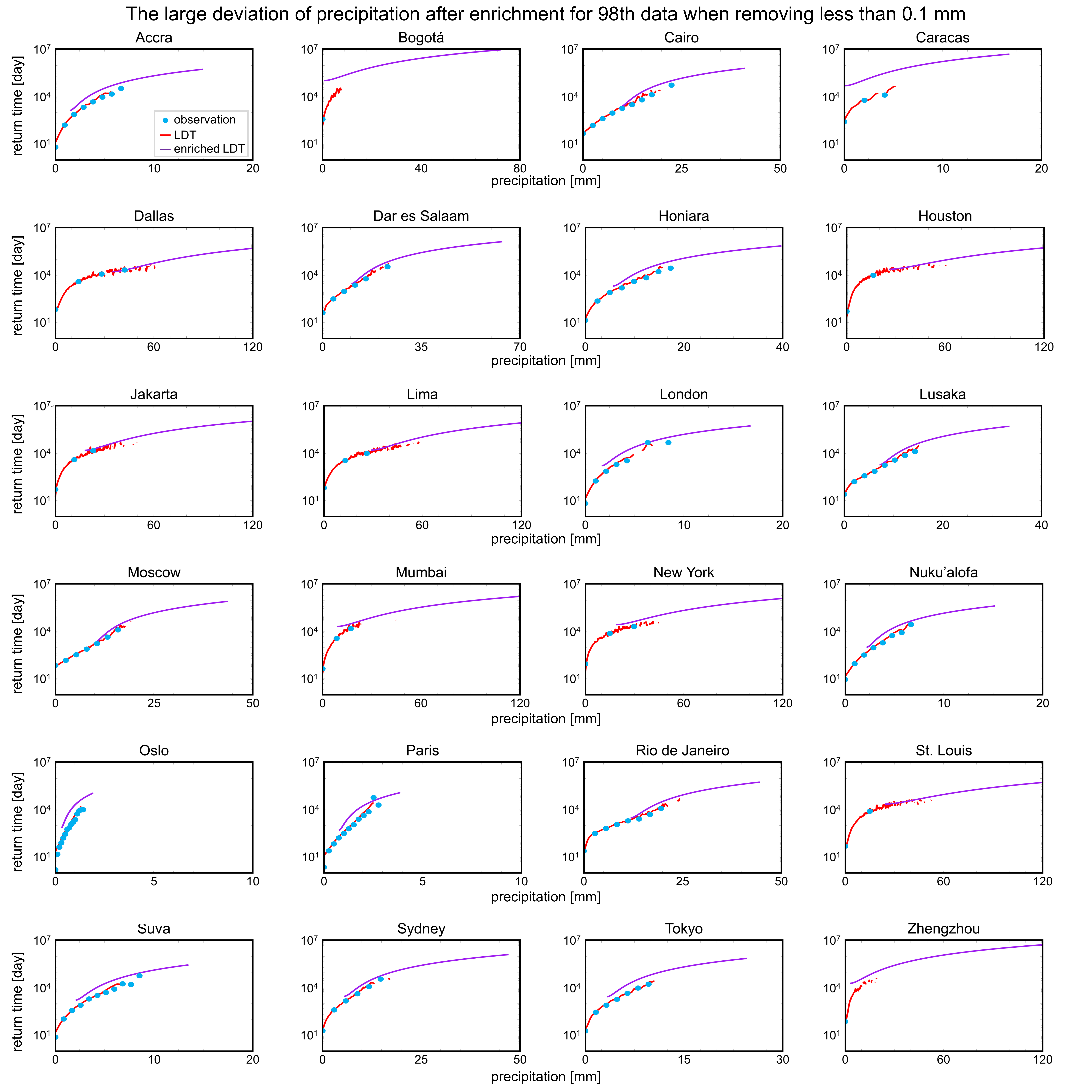}
    \caption{\textbf{Calculation of the large deviation of precipitation after enrichment by sampling from fits to 98th percentile distributions after removing less than 0.1 mm of precipitation days, as a supplement to Fig. ~3.} Historical precipitation data (blue points) and return time calculated using large deviation theory (LDT, red line) and enriched (purple line) by sampling at selected grid points, expanding the return time prediction range of the location. To reduce errors, the enriched return time were removed from the left side of the distribution peak.}
\end{figure}

\begin{figure}[H]
    \centering
    \includegraphics[width=\linewidth]{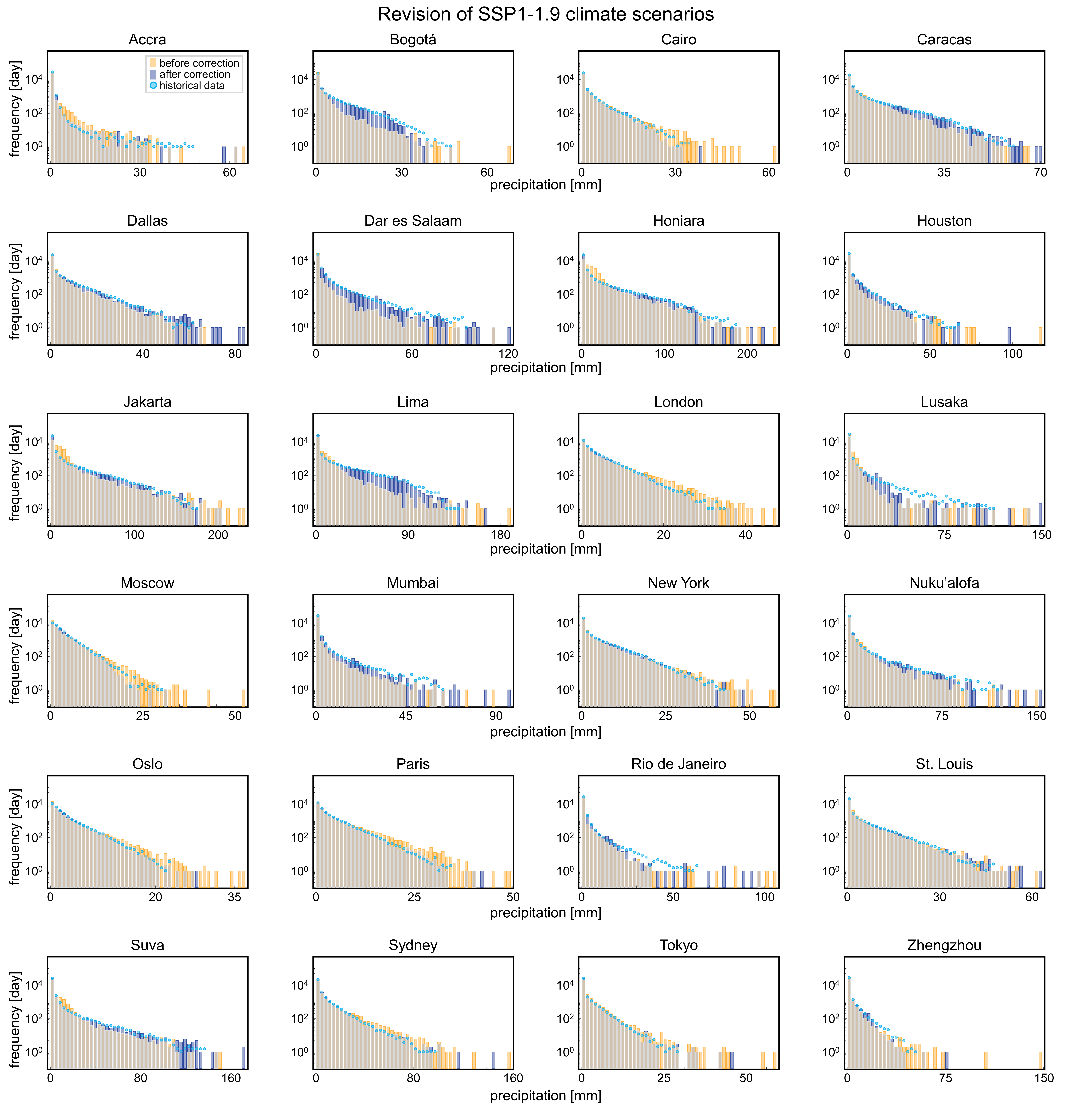}
    \caption{\textbf{Revision of SSP-1 climate scenarios, as a supplement to Fig. ~4.} Comparison of historical precipitation distributions before correction (yellow) and after correction (blue) with the SSP1-1.9 scenario distributions at selected grid points based on quantile mapping. It mainly increases the frequency of precipitation in the tail of the distribution.}
\end{figure}

\begin{figure}[H]
    \centering
    \includegraphics[width=\linewidth]{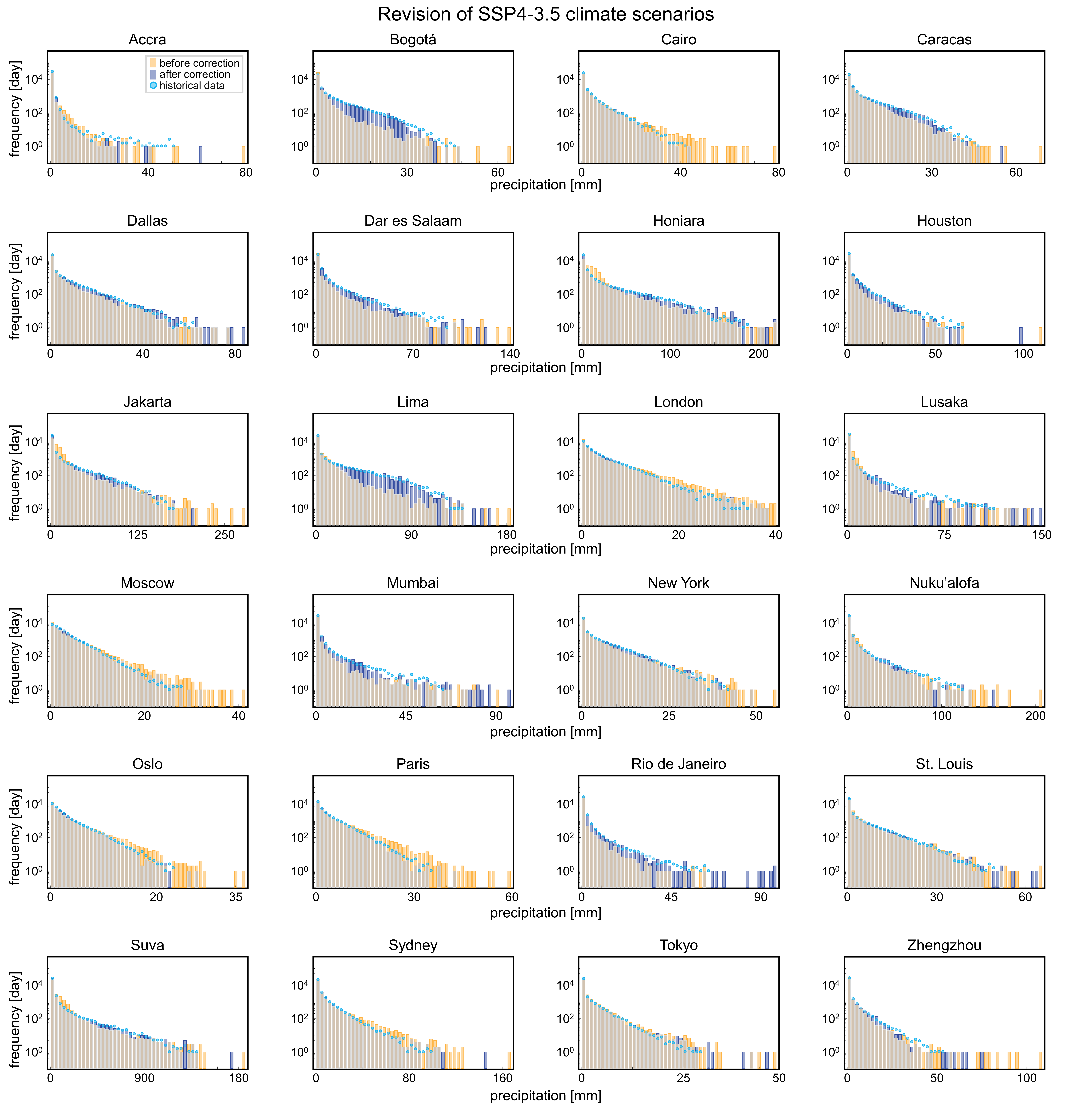}
    \caption{\textbf{Revision of SSP4-3.4 climate scenarios, as a supplement to Fig. ~4.} Comparison of historical precipitation distributions before correction (yellow) and after correction (blue) with the SSP4-3.4 scenario distributions at selected grid points based on quantile mapping. It mainly increases the frequency of precipitation in the tail of the distribution.}
\end{figure}

\begin{figure}[H]
    \centering
    \includegraphics[width=\linewidth]{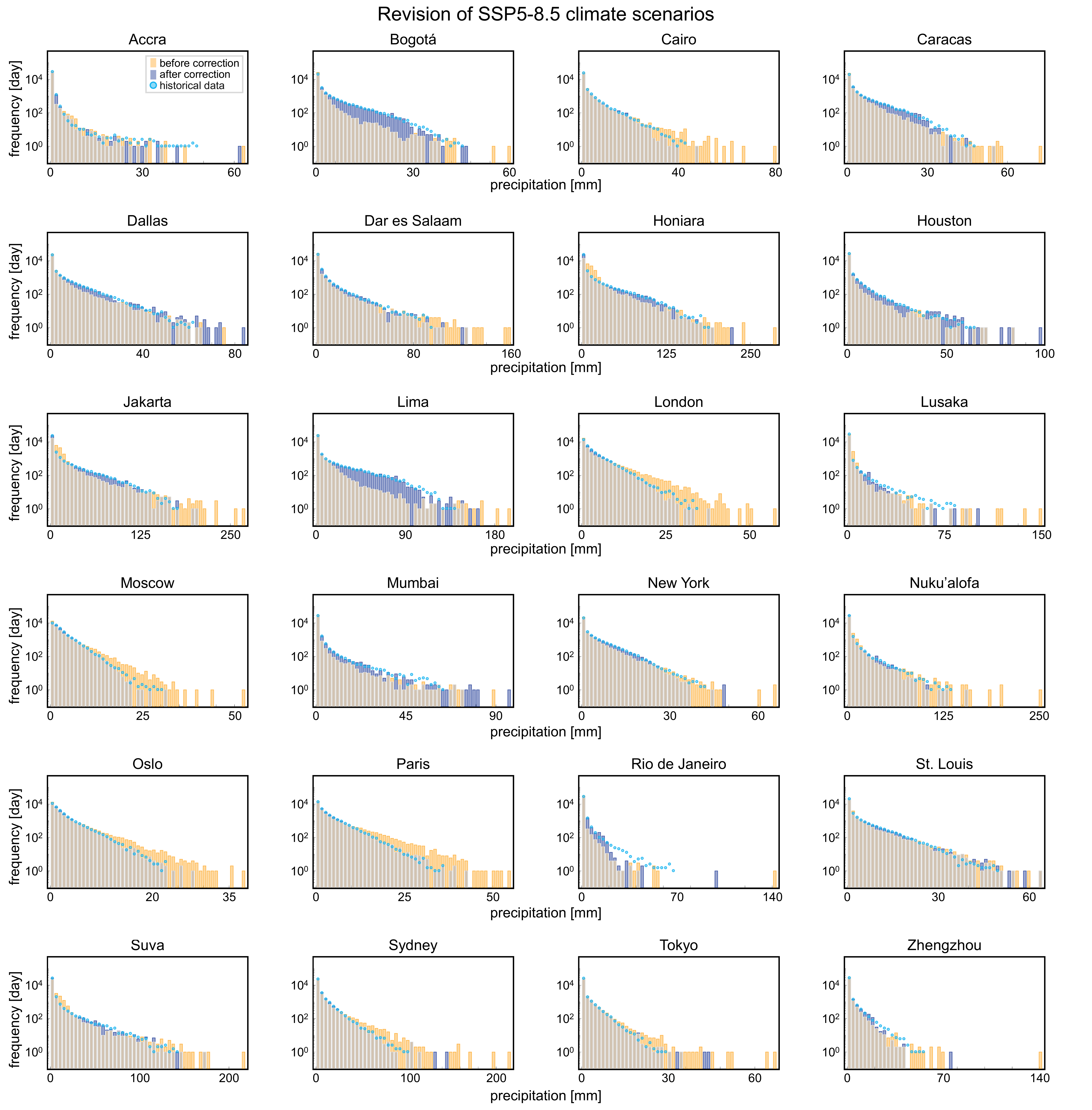}
    \caption{\textbf{Revision of SSP5-8.5 climate scenarios, as a supplement to Fig. ~4.} Comparison of historical precipitation distributions before correction (yellow) and after correction (blue) with the SSP4-3.4 scenario distributions at selected grid points based on quantile mapping. It mainly increases the frequency of precipitation in the tail of the distribution.}
\end{figure}

\begin{figure}[H]
    \centering
    \includegraphics[width=\linewidth]{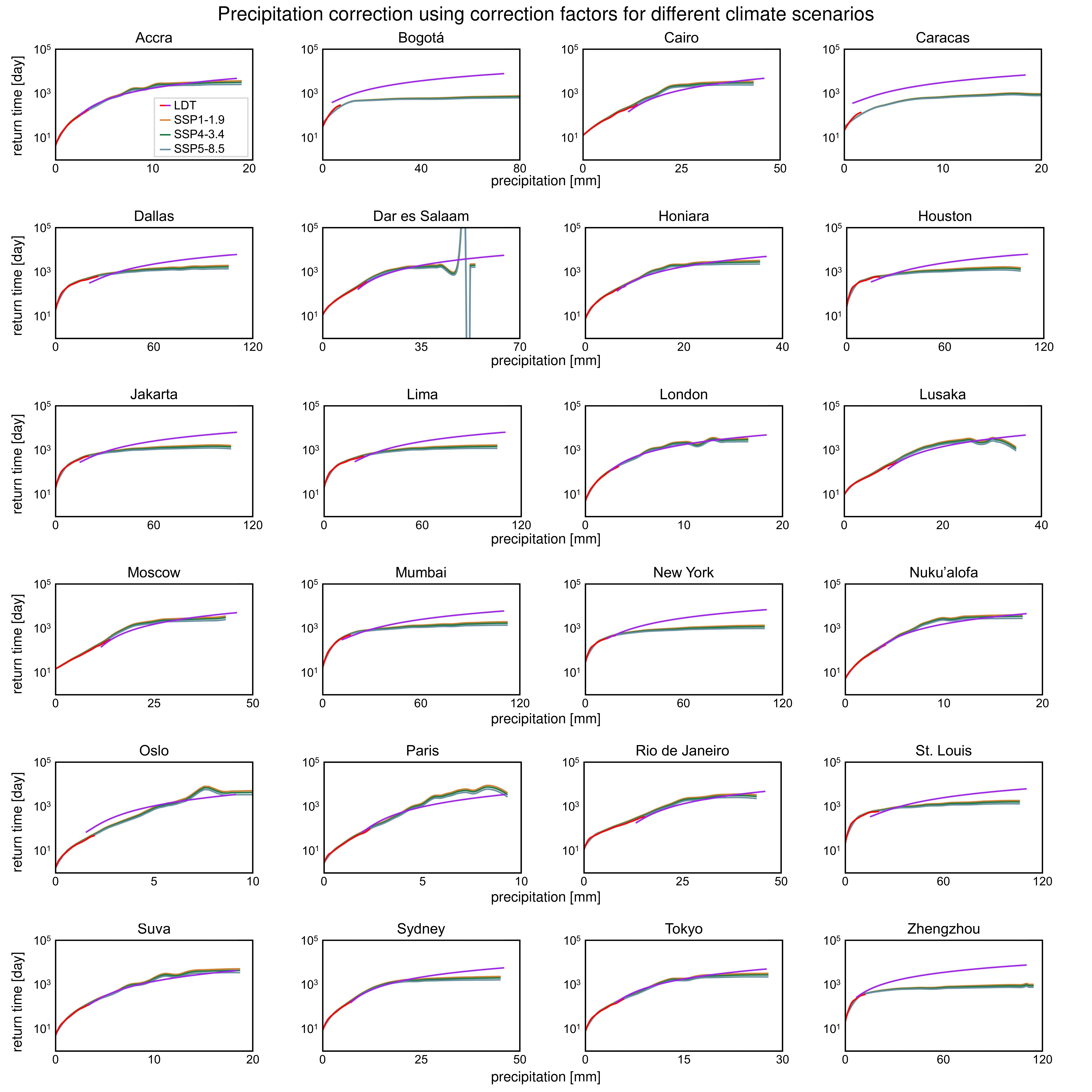}
    \caption{\textbf{Future precipitation return times for the 24 grid points after correction by the scenario specific correction factors, as a supplement to Fig.~4.} The large deviation analyses and complementary precipitation return time (red \& purple) and the corrected return time for the three climate scenarios SSP1-1.9 (yellow), SSP4-3.4 (green), and SSP5-8.5 (gray) are shown. At most locations, future precipitation generally has shorter return times, especially at larger precipitation amounts.}
\end{figure}

\begin{figure}[H]
    \centering
    \includegraphics[width=\linewidth]{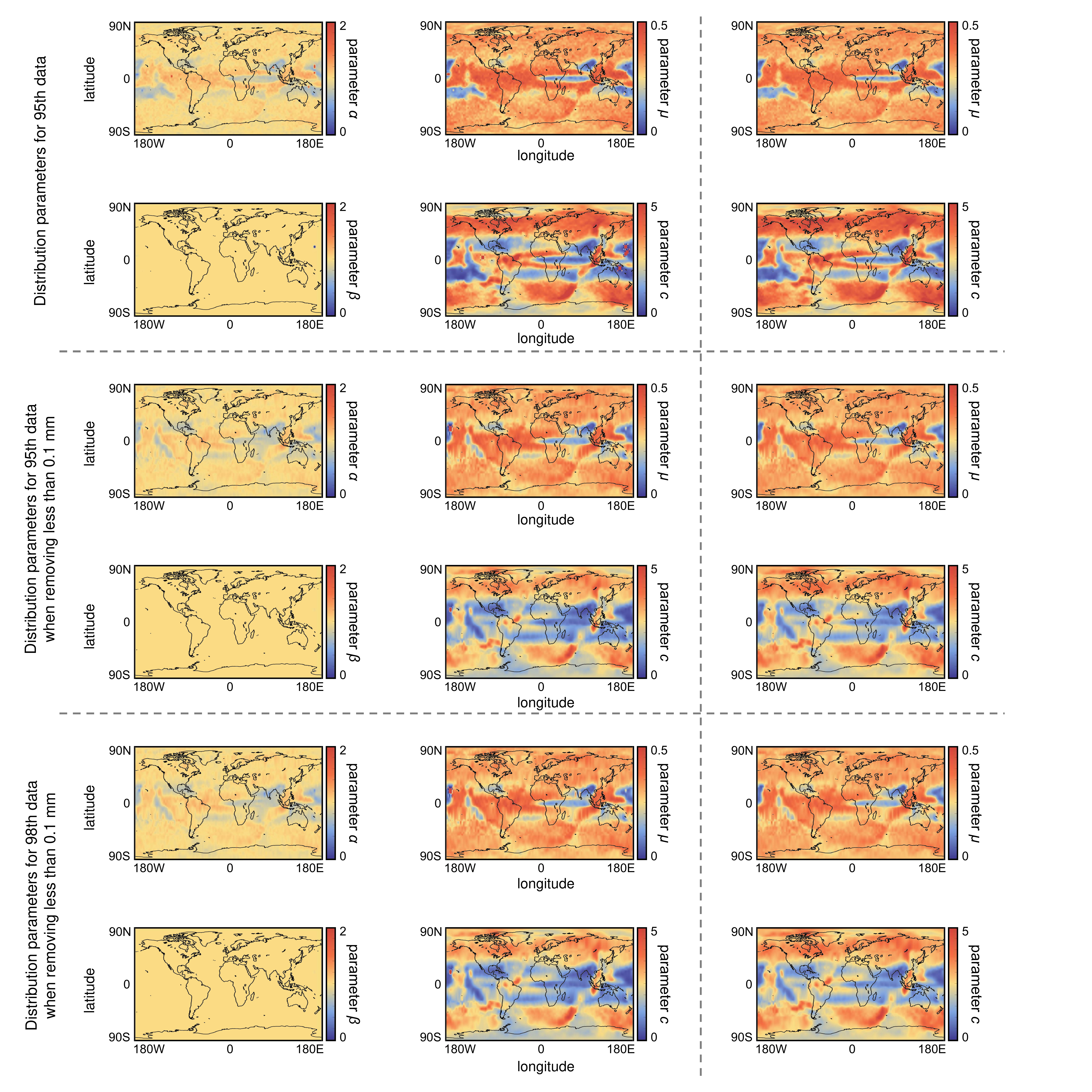}
    \caption{\textbf{Parameter laws under different data processing, as a supplement to Fig. ~6 in End Matter.} Shown are the spatial distributions of the four stable-distribution parameters $\alpha$, $\beta$, $\mu$, and $c$ and the two Landau-distribution parameters $\mu$ and $c$ for three cases: 95th percentile fitting with all precipitation days retained (top), 95th percentile fitting after removing days with precipitation below 0.1 mm (middle), and 98th percentile fitting after removing days with precipitation below 0.1 mm (bottom). The parameters $c$ and $\mu$ are typically lower at low latitudes and higher at high latitudes. Furthermore, the spatial distributions of $\mu$ and $c$ for the stable and Landau distributions are highly similar across the globe.}
\end{figure}

\begin{figure}[H]
    \centering
    \includegraphics[width=\linewidth]{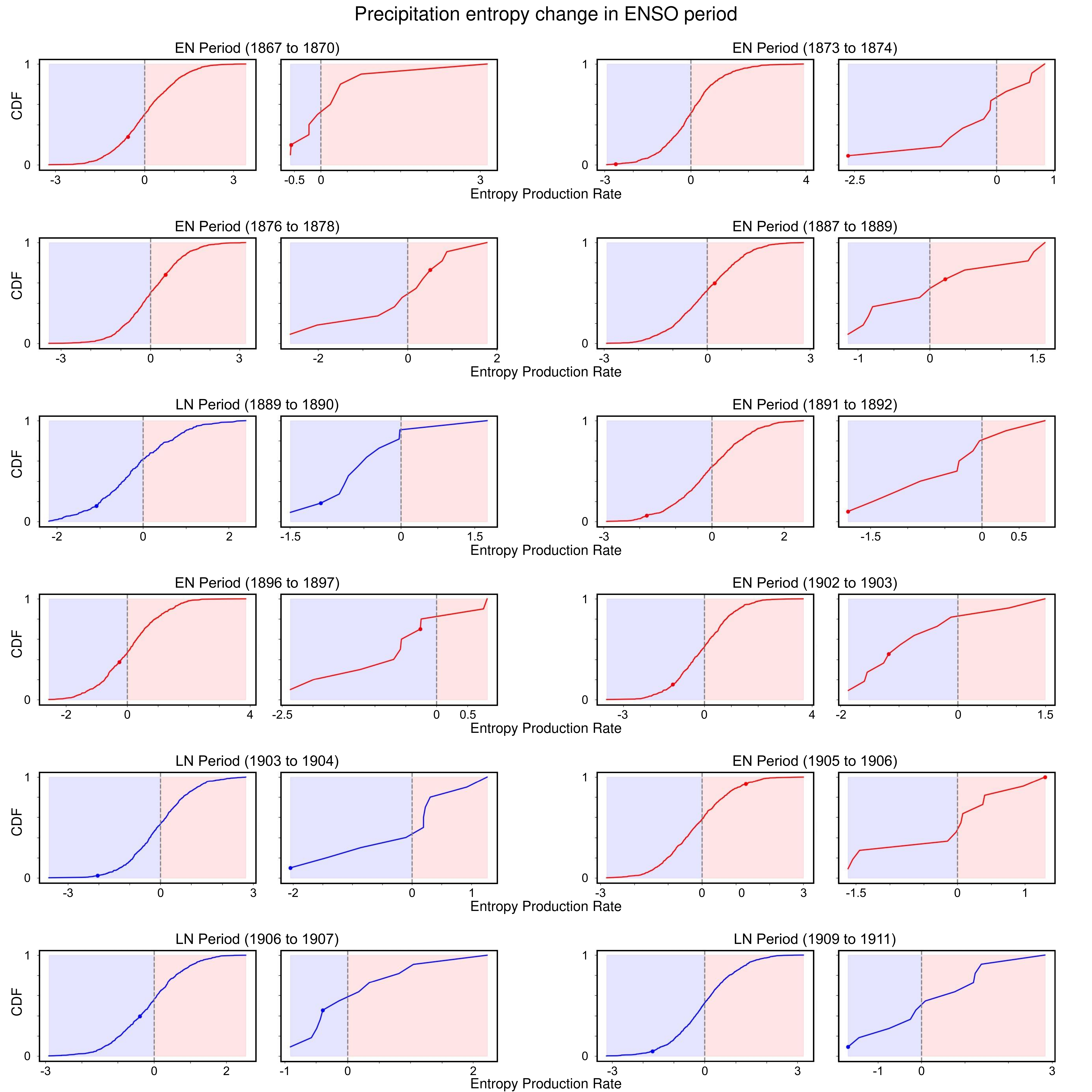}
    \caption{\textbf{The cumulative distribution functions of entropy production rates during ENSO transitions from 1900 to 2010, as supplemental to Section {Precipitation entropy changes during ENSO}.} Each sub-figure corresponds to a specific ENSO period. Results are presented for full trajectory periods (left) and extremes-only periods (right). El Ni\~{n}o (EN) and La Ni\~{n}a (LN) periods are represented by red and blue curves, respectively. The red-shaded region represents positive entropy production, while the blue-shaded region indicates negative entropy production. A dashed vertical line at zero entropy production serves as a reference for expected symmetry according to the fluctuation theorem. Significant deviations from symmetry in some periods are observed, particularly in extremes-only periods, where positive entropy production dominates.}
\end{figure}

\begin{figure}[H]
    \centering
    \includegraphics[width=\linewidth]{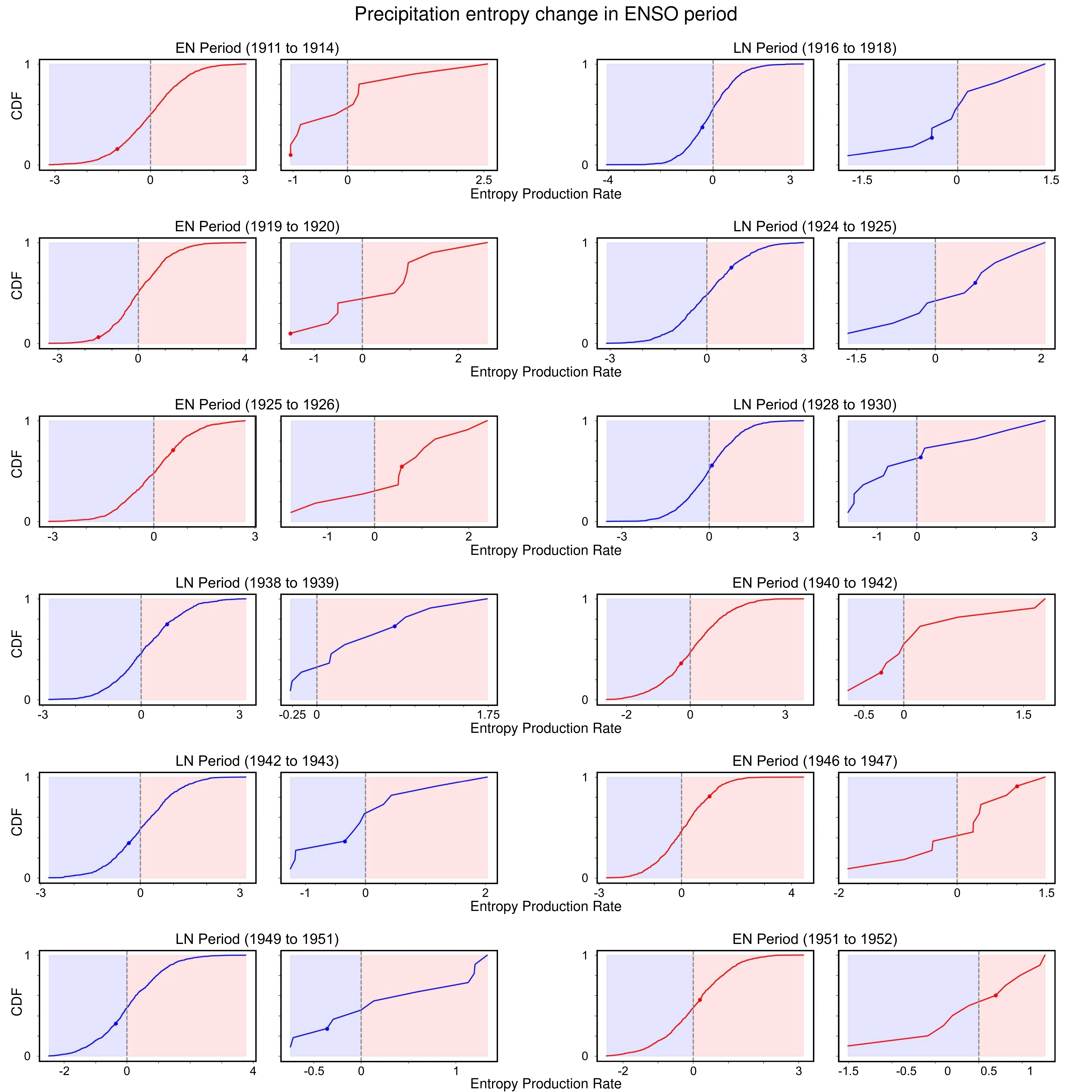}
    \caption{\textbf{The cumulative distribution functions of entropy production rates during ENSO transitions from 1911 to 1952, as supplemental to Section {Precipitation entropy changes during ENSO}.} Each sub-figure corresponds to a specific ENSO period. Results are presented for full trajectory periods (left) and extremes-only periods (right). El Ni\~{n}o (EN) and La Ni\~{n}a (LN) periods are represented by red and blue curves, respectively. The red-shaded region represents positive entropy production, while the blue-shaded region indicates negative entropy production. A dashed vertical line at zero entropy production serves as a reference for expected symmetry according to the fluctuation theorem. Significant deviations from symmetry in some periods are observed, particularly in extremes-only periods, where positive entropy production dominates.}
\end{figure}

\begin{figure}[H]
    \centering
    \includegraphics[width=\linewidth]{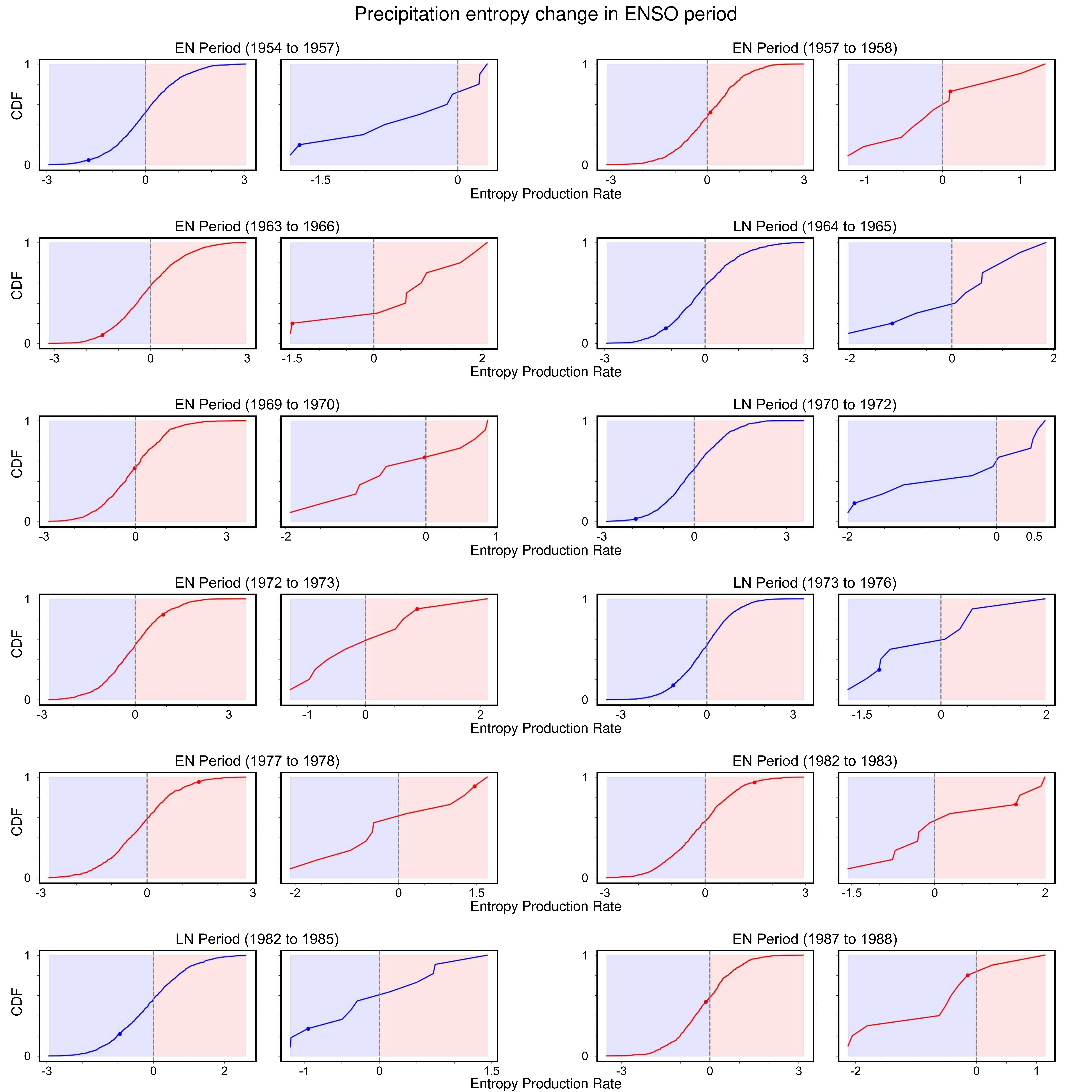}
    \caption{\textbf{The cumulative distribution functions of entropy production rates during ENSO transitions from 1952 to 1988, as supplemental to Section {Precipitation entropy changes during ENSO}.} Each sub-figure corresponds to a specific ENSO period. Results are presented for full trajectory periods (left) and extremes-only periods (right). El Ni\~{n}o (EN) and La Ni\~{n}a (LN) periods are represented by red and blue curves, respectively. The red-shaded region represents positive entropy production, while the blue-shaded region indicates negative entropy production. A dashed vertical line at zero entropy production serves as a reference for expected symmetry according to the fluctuation theorem. Significant deviations from symmetry in some periods are observed, particularly in extremes-only periods, where positive entropy production dominates.}
\end{figure}

\begin{figure}[H]
    \centering
    \includegraphics[width=\linewidth]{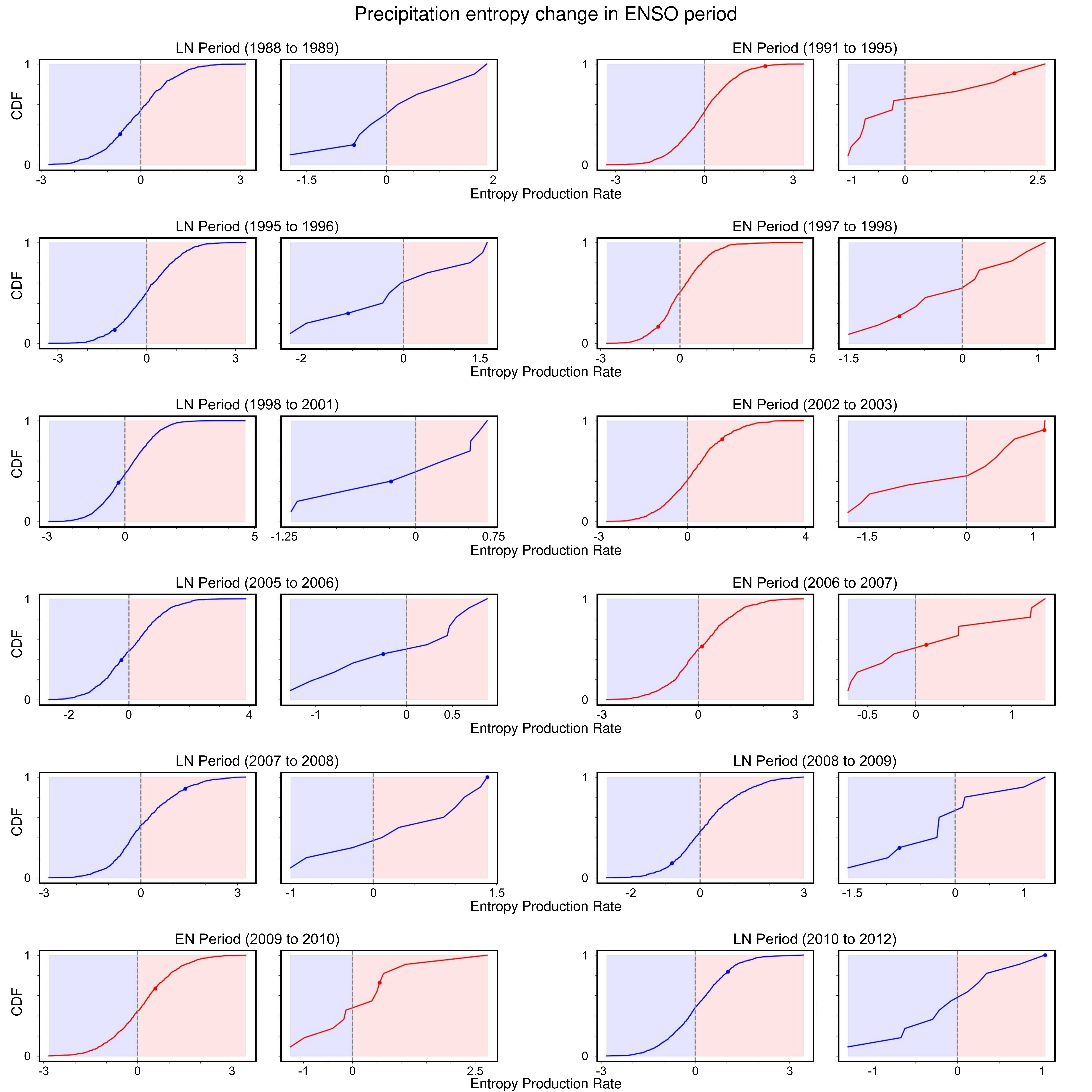}
    \caption{\textbf{The cumulative distribution functions of entropy production rates during ENSO transitions from 1988 to 2015, as supplemental to Section {Precipitation entropy changes during ENSO}.} Each sub-figure corresponds to a specific ENSO period. Results are presented for full trajectory periods (left) and extremes-only periods (right). El Ni\~{n}o (EN) and La Ni\~{n}a (LN) periods are represented by red and blue curves, respectively. The red-shaded region represents positive entropy production, while the blue-shaded region indicates negative entropy production. A dashed vertical line at zero entropy production serves as a reference for expected symmetry according to the fluctuation theorem. Significant deviations from symmetry in some periods are observed, particularly in extremes-only periods, where positive entropy production dominates.}
\end{figure}

\section*{Supplementary Tables}

\begin{table}[H]

\caption{\textbf{Predictions of future extreme precipitation events.} Twenty urbanized and populous cities were selected for analysis. For each city, the occurrence of extreme precipitation events was forecast and compared with the number of historical extreme precipitation events observed over the corresponding period. The analysis was conducted for thresholds of 40 mm, 70 mm, and 100 mm over 5-, 10-, 20-, and 30-year periods. The numerals in red denote historical extreme-precipitation events. For example, Beijing experienced an event exceeding 70 mm on July 20, 2016 ~\cite{LUO2020104895}; Guangzhou experienced an event exceeding 100 mm on May 22, 2020 ~\cite{zhang2024effect}; and Istanbul experienced an event exceeding 40 mm on September 7, 2009 ~\cite{komucscu2013analysis}. These historical events are consistent with the predictions.}
    %%\centering
\includegraphics[width=\linewidth]{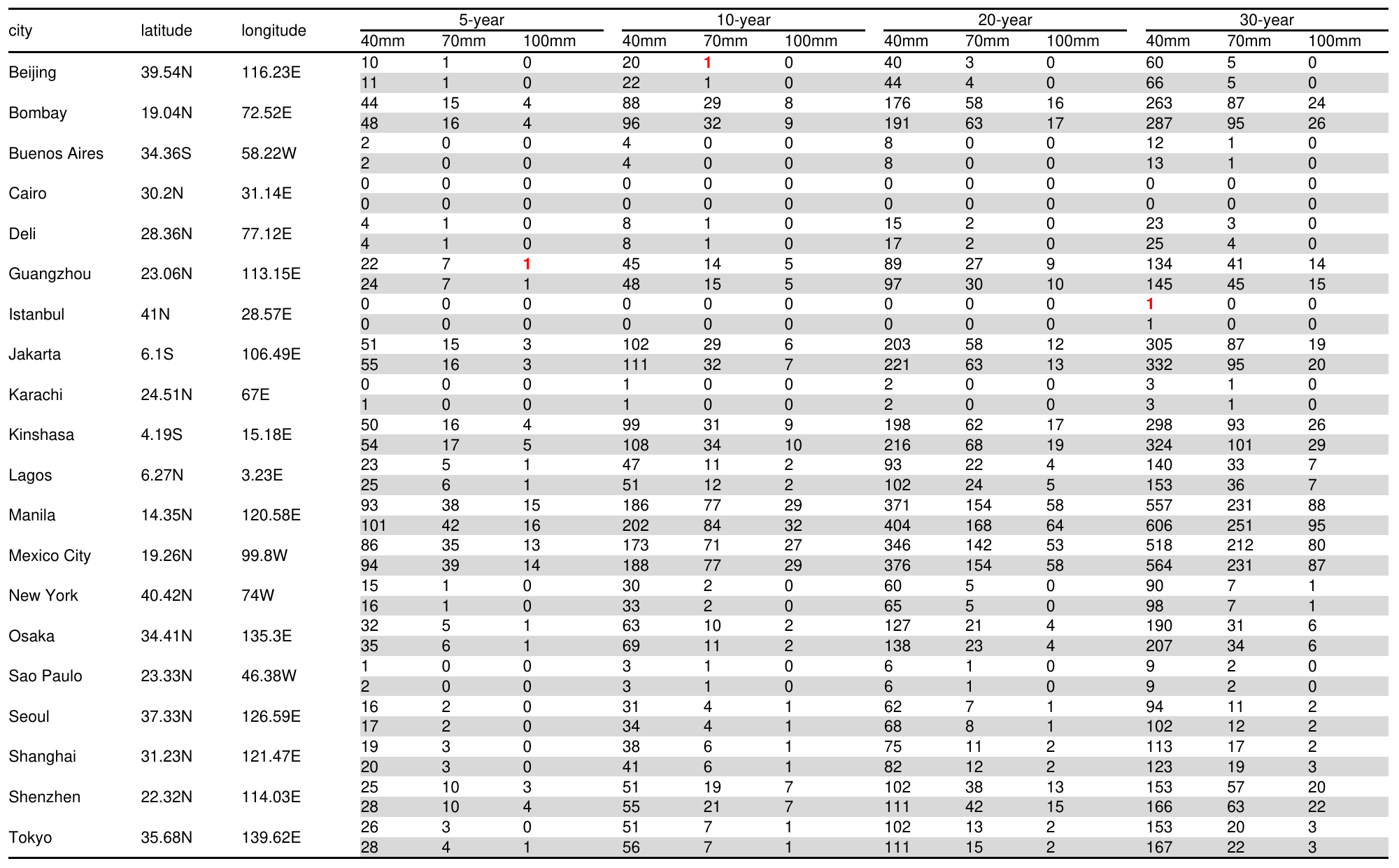}
\label{tab.1}
\end{table}

\begin{table}[H]  
\caption{\textbf{Precipitation statistics for different future climate scenarios.} Ten cities were selected for statistical analysis based on Table \ref{tab.1}, with the exclusion of cities that are geographically close or have less precipitation. Three climate scenarios were selected for the statistical analysis, SSP1-1.9 (top), SSP4-3.4 (middle), and SSP5-8.5 (bottom). The statistical analysis was conducted for thresholds of 40 mm, 70 mm, and 100 mm over 5-, 10-, 20-, and 30-year periods.}
%\centering
\includegraphics[width=\linewidth]{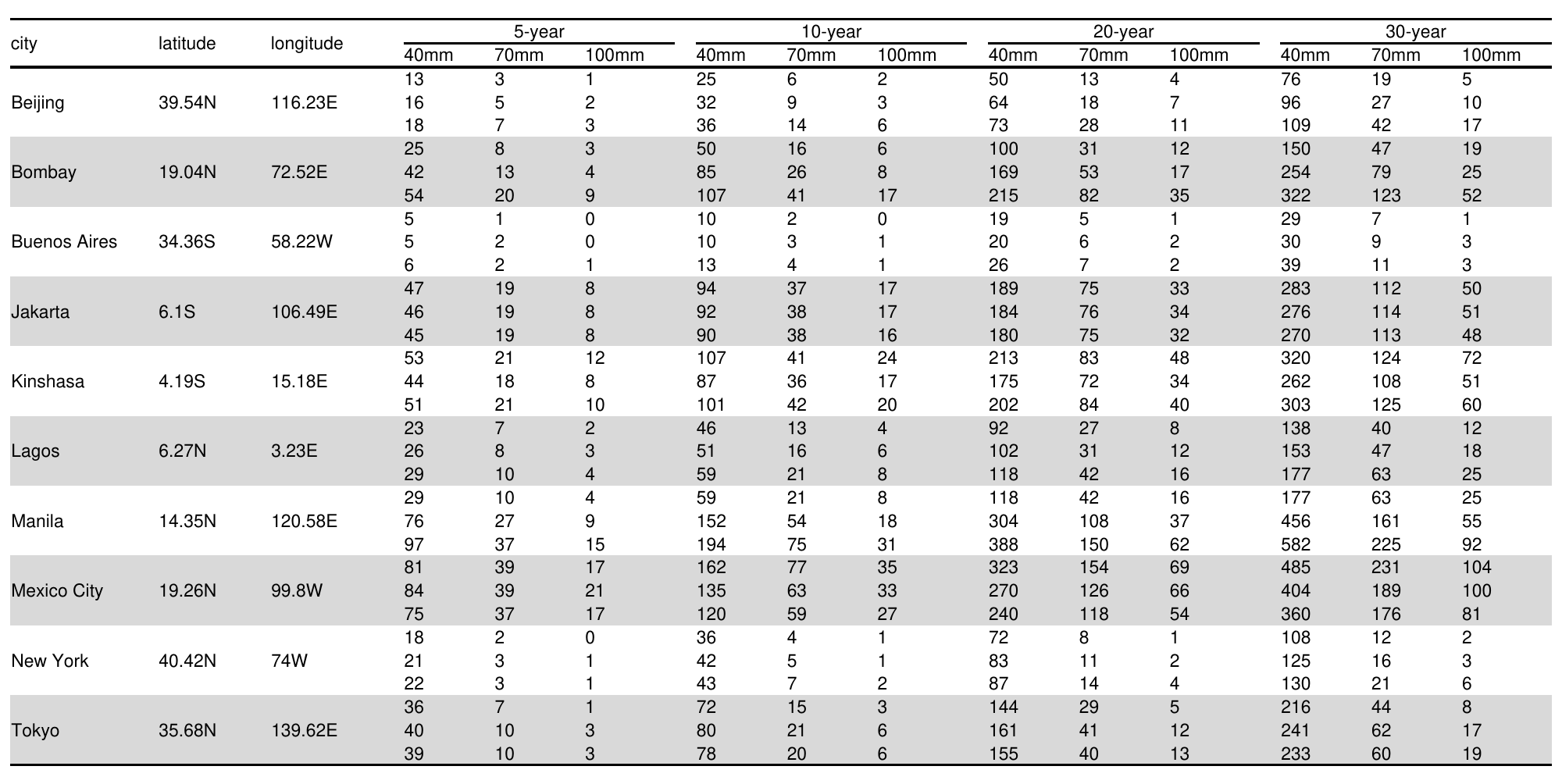}
\label{tab.2}
\end{table}

\putbib[reference]
\end{bibunit}

\end{document}